\documentclass[proof]{WileyASNA-v1}

\usepackage{color}

\articletype{Article Type}%

\received{}
\revised{}
\accepted{}

\begin{document}

\title{On the chemical abundance differences between 
the solar twin visual binary system 16 Cygni A and B}

\author[]{Yoichi Takeda}

\authormark{Y. TAKEDA}

\address[]{ 
\orgaddress{\state{11-2 Enomachi, Naka-ku, Hiroshima-shi, 730-0851}, \country{Japan}}}

\corres{\email{ytakeda@js2.so-net.ne.jp}}


\abstract{
The visual binary system 16~Cyg~A+B consists of similar solar twins, 
but a planetary companion is detected only in B.
An intensive spectroscopic differential analysis is carried out to the Sun, 16~Cyg~A, 
and 16~Cyg~B, with particular attentions being paid to (i) precisely establishing 
the differential atmospheric parameters/metallicity between A and B, and 
(ii) determining the important CNO abundances based on the lines of CH, NH, 
and OH molecules. The following results are obtained.  
(1) The Fe abundances (relative to the Sun) are [Fe/H]$^{\rm A} = +0.09$ and 
[Fe/H]$^{\rm B} =+0.06$ (i.e., A is slightly metal-rich than B by +0.03~dex). 
This lends support to the consequences of recently published papers, 
while the conclusion once derived by the author (almost the same metallicity 
for A and B) is acknowledged to be incorrect.
(2) The differential abundances ($\Delta$[X/H]) of volatile CNO with low $T_{\rm c}$ 
(condensation temperature) are apparently lower than those of refractory Fe group 
elements of higher $T_{\rm c}$, leading to a positive gradient in the 
$\Delta$[X/H] vs. $T_{\rm c}$ relation being more conspicuous for A than B.
This is qualitatively consistent with previous studies, though the derived 
slope is quantitatively somewhat steeper than that reported by other authors.
}


\maketitle


\section{Introduction}

The visual binary system 16~Cyg consists of two similar Sun-like 6th-magnitude stars
separated by $\sim 40$~arcsec: 16~Cyg~A (= HD~186408 = HR~7503; spectral type G1.5V) 
and 16~Cyg~B (= HD~186427 = HR~7504; G3V), the orbital elements of which are not yet 
well known because of the very long period (presumably $> 20000$~yr; cf. Kiselev \& Romanenko 
2011). This system is of particular astrophysical interest, since a planetary 
companion orbiting around B with a period of 800~d was found by Cochran et al. (1997)
by the radial velocity method (while any planet around A has never been reported so far).

Such a binary system in which only one component harbours a planet (while the other 
does not) may serve as a useful opportunity to investigate the impact of planet formation 
on the host star. That is, any difference in the chemical abundances between the two 
would provide us with valuable information on the star--planet connection (e.g., 
accretion of proto-planetary materials), since they should have originally formed 
from gas of the same composition.

Actually, several studies towards clarifying whether and how A and B show any chemical 
abundance differences were carried out from late 1990s to early 2000s; but the 
results were considerably diversified and no consensus could be accomplished
(see Table~11 in Takeda 2005).
I also conducted two decades ago precise differential analyses for selected 
8 solar-analogue stars (including 16~Cyg~A and B) by using a number of Fe lines 
in order to establish mutual parameter differences, and concluded that 
the metallicities of A and B are practically the same to a precision of $\sim 0.01$~dex
($[{\rm Fe/H}]^{\rm A} - [{\rm Fe/H}]^{\rm B} \simeq 0.00 \pm 0.01$).

However, several papers successively published after 2010 reported results 
against this conclusion; they all arrived at a similar consequence 
that 16~Cyg~A is slightly more metal-rich than B by $\sim$~0.03--0.05~dex, as summarised 
in Table~1. It was also argued in these recent studies (e,g., Tucci Maia et al. 2019) 
that not only the metallicity difference but also the different dependence of 
elemental abundances upon $T_{\rm c}$ (condensation temperature) exist between 
A and B, which may be related to the planet formation and evolution history of 
this binary system. 

\setcounter{table}{0}
\begin{table}
\begin{minipage}{80mm}
\caption{Recent results of atmospheric parameter differences between 16~Cyg~A and B.}
\begin{center}
\begin{tabular}{c@{ }c@{ }c@{ }c@{ }c}\hline
\hline
Reference & $\Delta T_{\rm eff}$ & $\Delta \log g$ & 
$\Delta v_{\rm t}$ & $\Delta A_{\rm Fe}$ \\
        & (K) & (dex) & (km~s$^{-1}$) & (dex) \\
\hline
Takeda (2005)            &  +39  &  $-0.07$ &  +0.08 &   0.00  \\
Schuler et al. (2011)    &  +43  &  $-0.02$ &  +0.10 &  +0.02$^{*}$  \\
Ram\'{\i}rez et al. (2011)    &  +64  &  $-0.05$ &  (0.0)$^{\dagger}$ &  +0.04$^{\ddagger}$  \\
Tucci Maia et al. (2014) &  +79  &  $-0.05$ &  +0.08 &  +0.05  \\
Nissen et al. (2017)     &  +56$^{\S}$  &  $-0.07$ &  +0.07 &  +0.03  \\
Tucci Maia et al. (2019) &  +69  &  $-0.05$ &  +0.08 &  +0.04  \\
Ryabchikova et al. (2022)&  +69  &  $-0.06$ &  +0.09 &  +0.03$^{\P}$  \\
\hline
\end{tabular}
\end{center}
Given in this table are the literature results on the differences (16~Cyg~A$-$16~Cyg~B) 
of $T_{\rm eff}$ (effective temperature), $\log g$ (logarithmic surface gravity), 
$v_{\rm t}$ (microturbulence), and $A_{\rm Fe}$ (Fe abundance) published after 2005.
See Table~11 of Takeda (2005) for those published before 2005.\\
$^{*}$Schuler et al. (2011) did not regard this difference as meaningful.\\ 
$^{\dagger}$This difference is zero because the same solar $v_{t}$ value of 
1.0 km~s$^{-1}$ was assumed for both A and B.\\
$^{\ddagger}$Mean of Fe~{\sc i} (+0.04~dex) and Fe~{\sc ii} (+0.03~dex).\\
$^{\S}$Mean of ionisation (+53~K) and excitation (+58~K) values.\\
$^{\P}$Mean of Fe~{\sc i} (+0.03~dex) and Fe~{\sc ii} (+0.02~dex).
\end{minipage}
\end{table}

This situation made me feel the necessity of revisiting this problem by using
observational data of sufficiently high-quality, because the equivalent width 
data employed in Takeda's (2005) analysis (taken from Takeda et al. 2005b) 
were measured from the spectra (published by Takeda et al. 2005a), which 
were not of satisfactory quality(signal-to-noise ratio was $\sim$~100--200) 
as viewed from present-day standard. 

Accordingly, I decided to conduct an intensive spectroscopic analysis 
(specifically targeted to 16~Cyg~A, 16~Cyg~B, and the Sun), in order to determine 
the mutual differences of atmospheric parameters and metallicity between these
three stars as precisely as possible (as done in Takeda 2005), where the equivalent 
widths are measured based on high-dispersion CFHT/ESPaDOnS spectra of sufficiently 
high quality in the public domain. Further, differential abundances of various 
elements are also derived by using such established atmospheric parameters to 
examine whether and how they show any meaningful dependence upon $T_{\rm c}$. 
This is the purpose of the present study. 

In the derivation of elemental abundances, special attention is paid to C, N, and O, 
which are the important volatile species of considerably low $T_{\rm c}$ ($< 200$~K) 
in comparison to the refractory species ($> 1000$~K) being in the majority. 
As a matter of fact, the slope of abundance vs. $T_{\rm c}$ relation is critically 
affected by these three light elements. However, the CNO abundances
of 16~Cyg~A and B derived in most of the recent studies (cf. Table~1) may not be 
very reliable because they were determined in most cases by neutral atomic 
lines which are generally few in number and weak in strength. In contrast, since 
differential CNO abundances of solar-analogue stars are well determinable from 
the numerous lines of hydride molecules (CH, NH, and OH) in the blue--UV region 
as very recently shown by Takeda (2023), the same molecular line features are 
again employed in this investigation. 
 
\section{Observational data}

\subsection{Adopted spectra}

The spectra of Vesta (Sun), 16~Cyg~A, and 16~Cyg~B, based on which 
the equivalent widths of spectral lines (to be used for determining the 
differential stellar parameters and chemical abundances) are measured, 
are the public-domain data originally obtained by Dr. J. Melend\'{e}z (P.I.) 
on 2013 March 4 (Vesta) and 2013 June 29 (16~Cyg~A and B) by the Echelle 
SpectroPolarimetric Device for Observation of Stars (ESPaDOnS) on the 3.6~m 
Canada--France--Hawaii Telescope (CFHT) at Mauna Kea. These are presumably
almost the same data as used by Tucci Maia et al. (2014).   
The following 7 files of reduced star-only spectra (resolution of $R = 81000$) 
were downloaded from web site of the Canadian Astronomy Data Centre:\footnote{ 
https://www.cadc.hia.nrc.gc.ca/AdvancedSearch/}
1611093i (Vesta; 1800~s exposure); 1634976i, 1634977i, and 1634978i (16~Cyg~A; 
each 280~s exposure); 1634979i, 1634980i, and 1634981i (16~Cyg~B: each 350~s exposure).
After co-adding the three spectra for 16~Cyg~A and 16~Cyg~B, the typical S/N ratios 
(at 6000~\AA) of the final spectra (covering 3700--8870~\AA) are estimated 
as $\sim 600$ (16~Cyg~A and B) and $\sim 400$ (Vesta).

\subsection{Measured lines and equivalent widths}

Takeda \& UeNo (2019) recently measured the equivalent widths of 565 lines 
(many are of Fe-group elements such as Ti, Cr, Fe, Ni) at 32 different points 
on the solar disk and studied their centre--limb variations in terms of the 
line properties. These lines were chosen as the basic set of candidate spectral 
lines to be measured in this study. 

Likewise, regarding the measurement of equivalent widths, the method of fitting with 
theoretical profile was employed as done in Takeda \& UeNo (2019):
(i) First, Takeda's (1995) spectrum-fitting technique is applied to an adequately 
chosen region comprising the relevant line, while varying the key parameters 
(elemental abundance $A$, macrobroadening parameter $v_{\rm M}$, and radial velocity 
$\Delta V_{\rm r}$) to accomplish their best-fit solutions. (ii) Then, based on 
such established abundance solution ($A$), the equivalent width of the line ($W_{\lambda}$) 
is inversely calculated.\footnote{The microturbulence ($v_{\rm t}$) and atmospheric 
model do not essentially matter in this case of $W_{\lambda}$ evaluation, as long as 
the same choice is made in both (i) and (ii). So, the solar model atmosphere with 
a fixed $v_{\rm t} = 1.0$~km~s$^{-1}$ was adopted.}
This method has a merit that there is no need to place the continuum level
in advance and that fitting is done with a more realistic simulated line profile
(in contrast to the conventional case of using Gaussian or Voigt profile fitting). 

In the measurement of $W_{\lambda}$, the spectral features around 
the target line  were carefully examined.
While showing the spectra of three stars (along with the theoretically synthesised 
solar spectrum) simultaneously on the computer display, I checked each line whether 
it is measurable (e.g., if judged to be unsuitable even in the spectrum of only one star, 
that line was discarded), and chose the fitting range (the same range is  
adopted for all three stars) by avoiding the contaminated part of the profile.
As a result, the equivalent widths of 531 lines (out of the 565 candidate lines)
could be measured after all, which are presented in the ``ewlines.dat''file 
of the supplementary material.

Such measured values of $W_{\lambda}^{\rm S}$ (Sun), $W_{\lambda}^{\rm A}$ (16~Cyg~A),
and $W_{\lambda}^{\rm B}$ (16~Cyg~B)\footnote{
In this article, Sun, 16~Cyg~A, and 16~Cyg~B are often represented for simplicity 
by one character of `S', `A', and `B', respectively.} for Fe~{\sc i} and Fe~{\sc ii} 
lines are compared with those adopted in Takeda (2005) (measured by the conventional 
Gaussian-fitting method; cf. Takeda et al. 2005b) in Fig.~1a, 1b, and 1c, respectively.
It can be seen from these figures that the $W_{2005}$ values used in Takeda (2005) 
tend to be systematically larger than the corresponding $W_{\lambda}$'s (newly 
measured in this study) by $\sim$~2--3\%, since the coefficients of linear-regression 
relations ($W_{2005} = a_{0} + a_{1}W_{\lambda}$) depicted in Figs.~1a--1c
by dashed lines are  ($a_{0}$, $a_{1}$) = (1.8~m\AA, 1.016), (2.1~m\AA, 1.015), 
(1.1~m\AA, 1.034), respectively. 

In Figs.~1d--1f are also shown the comparisons of the $W_{\lambda}$ values of 
Fe lines between the Sun, 16~Cyg~A, and 16~Cyg~B. Some characteristic trends 
can be read from these figures; e.g., the Fe lines tend to be stronger
in 16~Cyg~A/B than in the Sun (Figs.~1d and 1f), or $W_{\lambda}^{\rm A}$ 
(16~Cyg~A) is slightly larger than $W_{\lambda}^{\rm B}$ (16~Cyg~B) for 
Fe~{\sc ii} lines though no such difference is recognised for Fe~{\sc i} lines. 
These features are mainly due to the facts that 16~Cyg~A and B are more 
metal-rich than the Sun, and $\log g$ of 16~Cyg~A is lower than that of 16~Cyg~B.

\setcounter{figure}{0}
\begin{figure}
\begin{minipage}{80mm}
\includegraphics[width=8.0cm]{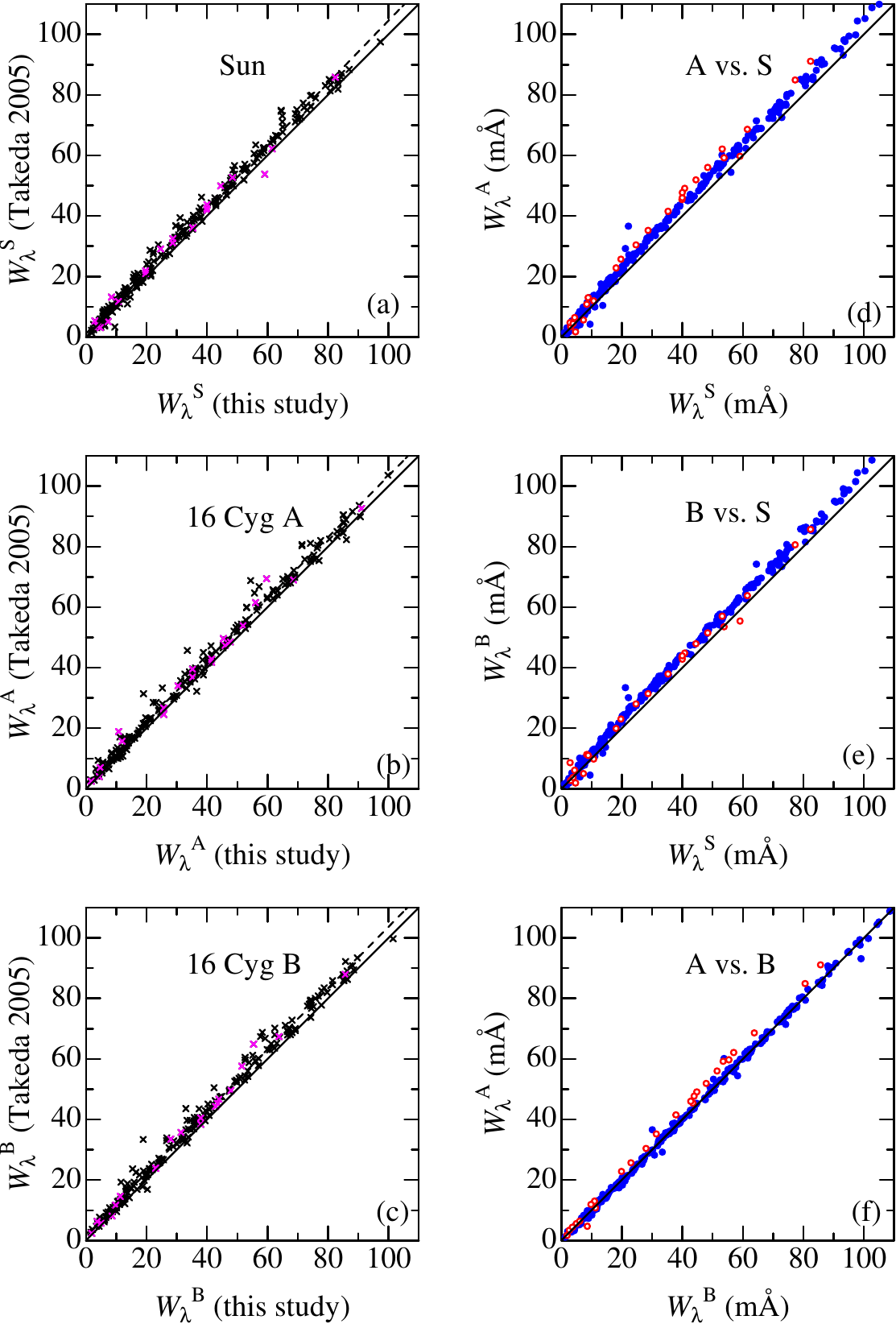}
\caption{Left panels (a-c): Comparison of the equivalent widths (in m\AA) 
of Fe lines adopted in this study (abscissa; derived by fitting with the 
theoretical spectrum) with those used in Takeda (2005)(ordinate; measured 
by the conventional Gaussian fitting method), where black crosses and pink 
crosses correspond to Fe~{\sc i} and Fe~{\sc ii} lines, respectively: 
(a) Sun, (b) 16~Cyg~A, and (c) 16~Cyg~B.
The linear-regression relation between $W_{\lambda}$(Takeda 2005) and 
$W_{\lambda}$(this study) is depicted by a dashed line (while the diagonal 
solid line is the guide for equal $W_{\lambda}$) in each panel. Right panels (d--f):  
Mutual comparisons of the equivalent widths of Fe lines used in this study, where 
blue filled symbols and red open symbols are for Fe~{\sc i} and Fe~{\sc ii} lines, 
respectively: (d) 16~Cyg~A vs. Sun, (e) 16~Cyg~B vs. Sun, and (f) 16~Cyg~A vs. 16~Cyg~B.
}
\label{fig1}
\end{minipage}
\end{figure}

\section{Differential analysis of atmospheric parameters}

\subsection{Principle of the method}

The method of differential analysis using Fe lines adopted in this study
is the same as that employed in Takeda (2005), which is essentially the
model atmosphere version of the traditionally well-known differential 
curve-of-growth analysis (e.g., Aller \& Greenstein 1960). This technique is quite 
effective in precisely establishing the differential abundances (simultaneously 
with the differences of atmospheric parameters) between two similar stars, because 
important error sources involved in the absolute analysis (e.g., $gf$ values) 
are cancelled out to accomplish a high precision. 

Let us denote the atmospheric parameters and the Fe abundance of the comparison 
star with a superscript ``0'' as ($T_{\rm eff}^{0}$, $\log g^{0}$, $v_{\rm t}^{0}$,
and $A^{0}$), which should be specified in advance but do not necessarily need to be precise.
Based on the equivalent widths $(W_{i}^{0}$, $i = 1, 2, \cdots, N)$,
the {\it empirical} $gf$ values $(\log gf_{i}^{0}, i= 1, \cdots ,N)$  are so 
derived (or adjusted) as to yield the same abundance $A^{0}$ for all lines. 

Next, the equivalent widths of the target star $(W_{i}$, $i = 1, 2, \cdots, N)$ 
are analysed by using these empirical $\log gf_{i}^{0}$ on a model atmosphere 
with parameters  $T_{\rm eff} (= T_{\rm eff}^{0} + \Delta T_{\rm eff})$, 
$\log g (= \log g^{0} + \Delta \log g)$, and 
$v_{\rm t} (= v_{\rm t}^{0} + \Delta v_{\rm t})$, 
to derive the abundance ($A_{i}$) and differential abundance 
$\Delta A_{i} (= A_{i} - A^{0})$ for each line.

This process is repeated with various combinations of 
($\Delta T_{\rm eff}$, $\Delta \log g$, $\Delta v_{\rm t}$).
Among many trials, the best solution 
($\Delta T_{\rm eff}^{*}$, $\Delta \log g^{*}$, $\Delta v_{\rm t}^{*}$)
should be regarded as the one yielding most consistent $(\Delta A_{i}, i = 1, 2, \cdots, N)$ 
without showing any systematic dependence upon the line properties.

More precisely, it is required that the following three conditions should be met.
\begin{itemize}
\item
The abundances ($\Delta A_{i,1}$) of Fe~{\sc i} lines do not depend upon 
the line strengths ($W_{i,1}$) (curve-of-growth matching).
\item
The abundances ($\Delta A_{i,1}$) of Fe~{\sc i} lines do not depend upon 
the lower excitation potential $\chi_{\rm low}$ (excitation equilibrium), 
\item
The mean abundances of Fe~{\sc i} lines ($\Delta A_{1}$)
and Fe~{\sc ii} lines ($\Delta A_{2}$) are equal to each other 
(ionisation equilibrium).
\end{itemize}

The desired solutions ($\Delta T_{\rm eff}^{*}$, $\Delta \log g^{*}$,  
$\Delta v_{\rm t}^{*}$) matching these three conditions are determined
in the same manner as adopted in Takeda (2005) (see Takeda et al. 
2002 for more details). 

The quantities $\Delta A_{1}$ (mean of Fe~{\sc i} 
abundances), $\Delta A_{2}$ (mean of Fe~{\sc ii} abundances),
$\sigma_{1}$ (standard deviation of Fe~{\sc i} abundances), and
$\sigma_{2}$ (standard deviation of Fe~{\sc ii} abundances) are 
calculated for any given set of $(\Delta A_{i}, i=1, 2, \cdots, N)$ as follows.
\begin{equation}
 \Delta A_{1} \equiv \sum_{i=1}^{N_{1}} \Delta A_{i,1} / N_{1},
\end{equation}
\begin{equation}
 \Delta A_{2} \equiv \sum_{i=1}^{N_{2}} \Delta A_{i,2} / N_{2},
\end{equation}
\begin{equation}
\sigma_{1} \equiv \sqrt{ 
\sum_{i=1}^{N_{1}}{(\Delta A_{i,1} -  \Delta A_{1})^{2}} \left. \middle/ N_{1}\right.},
\end{equation}
and
\begin{equation}
\sigma_{2} \equiv \sqrt{ 
\sum_{i=1}^{N_{2}}{(\Delta A_{i,2} -  \Delta A_{2})^{2}} \left. \middle/ N_{2}\right.},
\end{equation}
where $N_{1}$ and $N_{2}$ are the numbers of adopted Fe~{\sc i} and Fe~{\sc ii} lines, respectively.
Further, a function ($D$) is defined as 
\begin{equation}
D^{2} \equiv \sigma_{1}^{2} + (\Delta A_{1} - \Delta A_{2})^{2}, 
\end{equation}
which may be called as ``dispersion function''.\footnote{Here, $\sigma_{2}$ is not included in $D$. 
Actually, there is no merit in doing so (which makes the functional property of $D$ even complex), 
because the number of Fe~{\sc ii} lines is considerably smaller than that of Fe~{\sc i} lines 
and the parameter sensitivity is totally different.}
Then the point accomplishing the minimum $D$ in the 3-dimensional space 
($\Delta T_{\rm eff}$, $\Delta \log g$, $\Delta v_{\rm t}$) yields the solutions of
differential parameters, from which the differential abundance 
$\Delta A$ (= $\Delta A_{1}$ = $\Delta A_{2}$)
is also established.   

\subsection{Practical procedures and results}

Differential analyses of atmospheric parameters in this study are applied to three
pairs (target star vs. comparison star): 16~Cyg~A vs. Sun, 16~Cyg~B vs. Sun, and 
16~Cyg~A vs. 16~Cyg~B. The following procedures are adopted for this purpose. 

First, many sets of abundances ($A_{i}, i=1, 2, \cdots, N$)\footnote{
These $A_{i}$ values are ``absolute'' abundances derived by using 
Kurucz \& Bell's (1995) $gf$ values (as given in ``ewlines.dat'' of supplementary 
material), which however are eventually cancelled out and irrelevant in 
``differential'' abundances ($\Delta A_{i}$) under question.} 
for all Fe lines were calculated from the observed equivalent widths of the target star 
($W_{i}, i=1, 2, \cdots, N$) by using Kurucz's (1993) WIDTH9 program 
for an extensive grid of 41$\times$21$\times$21$\times$2 combinations of parameters: 
41 $T_{\rm eff}$ (from 5700 to 5900~K with a step of 5~K), 21 $\log g$ (from 4.20 
to 4.60 with a step of 0.02~dex), 21 $v_{\rm t}$ (from 0.80 to 1.20~km~s$^{-1}$ 
with a step of 0.02~km~$^{-1}$), and 2 metallicities [M/H] (0.0 and +0.1). 
The necessary model atmospheres were generated by interpolating Kurucz's (1993) 
grid of ATLAS9 models.

Then, by intentionally choosing $T_{\rm eff}^{0}$, $\log g^{0}$, and $v_{\rm t}^{0}$ 
of the comparison star to coincide with any node of the $A_{i}$ grid (i.e., 
multiples of 5~K, 0.02~dex, and 0.02~km~s$^{-1}$, respectively; cf. Table~2), 
similar grid of differential abundances ($\Delta A_{i} = A_{i} -A^{0}$) can be constructed, 
from which $\Delta A_{1}$, $\Delta A_{2}$, $\sigma_{1}$, and $D$ are further defined 
in the 3-dimensional parameter space ($\Delta T_{\rm eff}$, $\Delta \log g$, 
$\Delta v_{\rm t}$).\footnote{Since the effect of model metallicity ([M/H] = [Fe/H]) 
on the abundance ($A$) is not significant (cf. Sect.~3.4), it was treated in an approximate
manner: Two sets of abundance grids ($A_{i}^{0.0}$ and $A_{i}^{0.1}$ 
corresponding to [M/H]~=~0.0 and 0.1) are prepared, and they are linearly interpolated 
in terms of given $A_{\rm give}$  as $A_{i}$ = $A_{i}^{0.0}$ + 
$(A_{i}^{0.1}- A_{i}^{0.0})(A_{\rm give} - 7.50)/(7.60-7.50)$, where $A_{{\rm Fe},\circ} = 7.50$
is the solar Fe abundance (in the usual normalisation of $A_{\rm H} = 12$). 
Regarding the choice of $A_{\rm give}$, it is naturally $A_{0}$ for the comparison star, 
while a reasonable value is assumed for the target star, the validity of which 
is checked after the final solution has been obtained.
} 

As in Takeda (2005), the Fe lines to be used for calculating $\Delta A_{1}$, 
$\Delta A_{2}$, $\sigma_{1}$, and $D$ were restricted to those not stronger than 100~m\AA\  
(in terms of the solar $W_{\lambda}$). Besides, those lines appreciably deviating from 
the mean trend (judged by Chauvenet's criterion) were discarded.
The lines finally adopted in each differential analysis are indicated
in ``relabunds\_AtoS.dat'', ``relabunds\_BtoS.dat'', and ``relabunds\_AtoB.dat''of 
the supplementary material (see column 82 therein).

Since the necessary quantities ($\Delta A_{1}$, $\Delta A_{2}$, $\sigma_{1}$, and $D$) 
have been evaluated as functions of ($\Delta T_{\rm eff}$, $\Delta \log g$, 
$\Delta v_{\rm t}$), the solutions of these differential parameters are determinable.
Although Takeda (2005) directly obtained the solution of minimum $D$ by applying 
the numerical optimisation algorithm (downhill simplex method; cf. Takeda et al. 2002),
somewhat different manual approaches are tried here, in order to 
clarify the functional behaviour of $D$ in the parameter space.

At each fixed $\Delta v_{\rm t}$, the location minimising $D$ is searched in the 
$(\Delta T_{\rm eff}, \Delta \log g)$ plane. The quantities at this local minimum 
are denoted with a dagger, such as $D^{\dagger}$, $\Delta T_{\rm eff}^{\dagger}$, 
and $\Delta \log g^{\dagger}$. Then, the minimum of the 1-dimensional function 
$D^{\dagger}(\Delta v_{\rm t})$ yields the desired solution $\Delta v_{\rm t}^{*}$ 
(indicated by an asterisk). In this way, the final solutions of differential 
parameters $\Delta T_{\rm eff}^{*}$, $\Delta \log g^{*}$, and $\Delta v_{\rm t}^{*}$ 
(along with the corresponding $\Delta A^{*}$) are established.
Application of this procedure to the three cases is graphically illustrated in Fig.~2.  
The results are summarised in Table~2.

\setcounter{figure}{1}
\begin{figure}
\begin{minipage}{80mm}
\includegraphics[width=8.0cm]{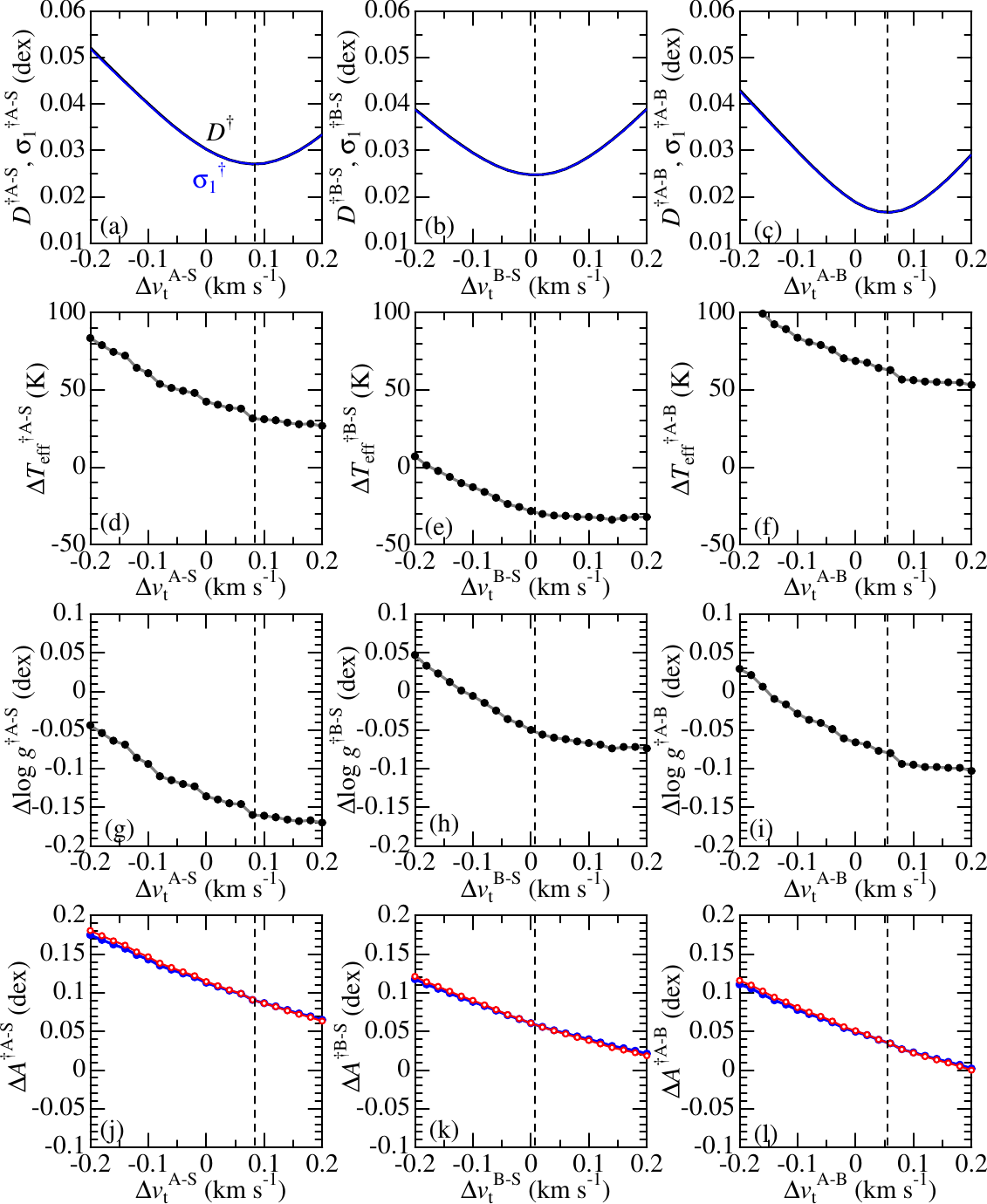}
\caption{
Top-row panels (a--c): $D^{\dagger}(\Delta T_{\rm eff}^{\dagger},\Delta\log g^{\dagger})$ 
and $\sigma_{1}^{\dagger}(\Delta T_{\rm eff}^{\dagger},\Delta\log g^{\dagger})$ plotted against 
$\Delta v_{\rm t}$, where $(\Delta T_{\rm eff}^{\dagger},\Delta\log g^{\dagger})$
is the (pseudo-)minimum position of $D$ for each given $\Delta v_{\rm t}$ 
($D$ and $\sigma_{1}$ are almost indiscernible here).
2nd-row panels (d--f): Run of $\Delta T_{\rm eff}^{\dagger}$ with $\Delta v_{\rm t}$.
3rd-row panels (g--i): Run of $\Delta \log g^{\dagger}$ with $\Delta v_{\rm t}$.
Bottom-row panels (j--l): Run of $\Delta A_{1}^{\dagger}$ (blue filled symbols) and 
$\Delta A_{2}^{\dagger}$ (red open symbols) with $\Delta v_{\rm t}$.
In each panel, the position of final $\Delta v_{\rm t}^{*}$ solution is indicated 
by the vertical dashed line. 
The left, centre, and right panels are for 16~Cyg~A $-$ Sun, 
16~Cyg~B $-$ Sun, and 16~Cyg~A $-$ 16~Cyg~B, respectively.
}
\label{fig2}
\end{minipage}
\end{figure}

\setcounter{table}{1}
\begin{table*}
\begin{minipage}{150mm}
\caption{Resulting solutions of differential atmospheric parameters.}
\begin{center}
\begin{tabular}{ccccccccc}\hline
\hline
target$-$[comparison] & [$T_{\rm eff}^{0}$] & $\Delta T_{\rm eff}$ &
                [$\log g^{0}$] & $\Delta \log g$ &
                [$v_{\rm t}^{0}$] & $\Delta v_{\rm t}$ &
                [$A_{\rm Fe}^{0}$] & $\Delta A_{\rm Fe}$ \\
   & (K)  & (K)  & (dex) & (dex) & (km~s$^{-1}$) & (km~s$^{-1}$) & (dex) & (dex)\\
\hline
16~Cyg~A$-$[Sun] & [5780] & +31.4 & [4.44] & $-0.160$ & [1.00] & +0.083 & [7.50] & +0.090 \\
               &  & ($\pm 12.1$) &  & ($\pm 0.028$) & & ($\pm 0.018$) &  &  ($\pm 0.009$) \\ 
16~Cyg~B$-$[Sun] & [5780] & $-29.2$ & [4.44] & $-0.052$ & [1.00] & +0.007 & [7.50] & +0.059 \\
               &  & ($\pm 13.4$) &  & ($\pm 0.033$) & & ($\pm 0.022$) &  &  ($\pm 0.009$) \\ 
16~Cyg~A$-$[16~Cyg~B] & [5750] & +63.1 & [4.38] & $-0.079$ & [1.00] & +0.055 & [7.56] & +0.036 \\
               &  & ($\pm 8.9$) &  & ($\pm 0.021$) & & ($\pm 0.012$) &  &  ($\pm 0.006$) \\ 
\hline
\end{tabular}
\end{center}
Values in the bracket are the assumed parameters of the comparison star ($p^{0}$), while
the signed values are the solutions of the relative parameter difference 
($\Delta p$; whose errors are given with $\pm$ in the next row; cf. Sect. 3.3).
Regarding the adopted parameters of [16~Cyg~B] as a comparison star, they were derived
from the results of 16~Cyg~B$-$[Sun] analysis as $5780-29.2 (\rightarrow 5750)$~K, 
$4.44-0.052 (\rightarrow 4.38)$~dex, $1.0+0.007 (\rightarrow 1.00)$~km~s$^{-1}$, 
and $7.50+ 0.059 (\rightarrow 7.56)$. 
\end{minipage}
\end{table*}

These solutions were also checked by using an alternative approach focusing on 
$\sigma_{1}$, $\Delta A_{1}$, and $\Delta A_{2}$, separately (rather than $D$ itself).
At each fixed $\Delta \log g$, the location minimising $\sigma_{1}$ is searched in the 
$(\Delta T_{\rm eff}, \Delta v_{\rm t})$ plane. The quantities at this local minimum 
are denoted as $\sigma_{1}^{\dagger}$, $\Delta T_{\rm eff}^{\dagger}$, 
$\Delta v_{\rm t}^{\dagger}$, $\Delta A_{1}^{\dagger}$ and $\Delta A_{2}^{\dagger}$. 
Then, making use of the fact that $\Delta A_{1}$ is inert but $\Delta A_{2}$ is sensitive
to a change in $\Delta\log g$, the intersection of  $\Delta A_{1}^{\dagger}(\Delta \log g)$
and $\Delta A_{2}^{\dagger}(\Delta \log g)$ yields the solution of $\Delta \log g^{*}$,
from which $\Delta T_{\rm eff}^{*}$, $\Delta v_{\rm t}^{*}$, and  $\Delta A^{*}$ are 
further obtained. Fig.~3 displays how this way of solution search actually worked.  

\setcounter{figure}{2}
\begin{figure}
\begin{minipage}{80mm}
\includegraphics[width=8.0cm]{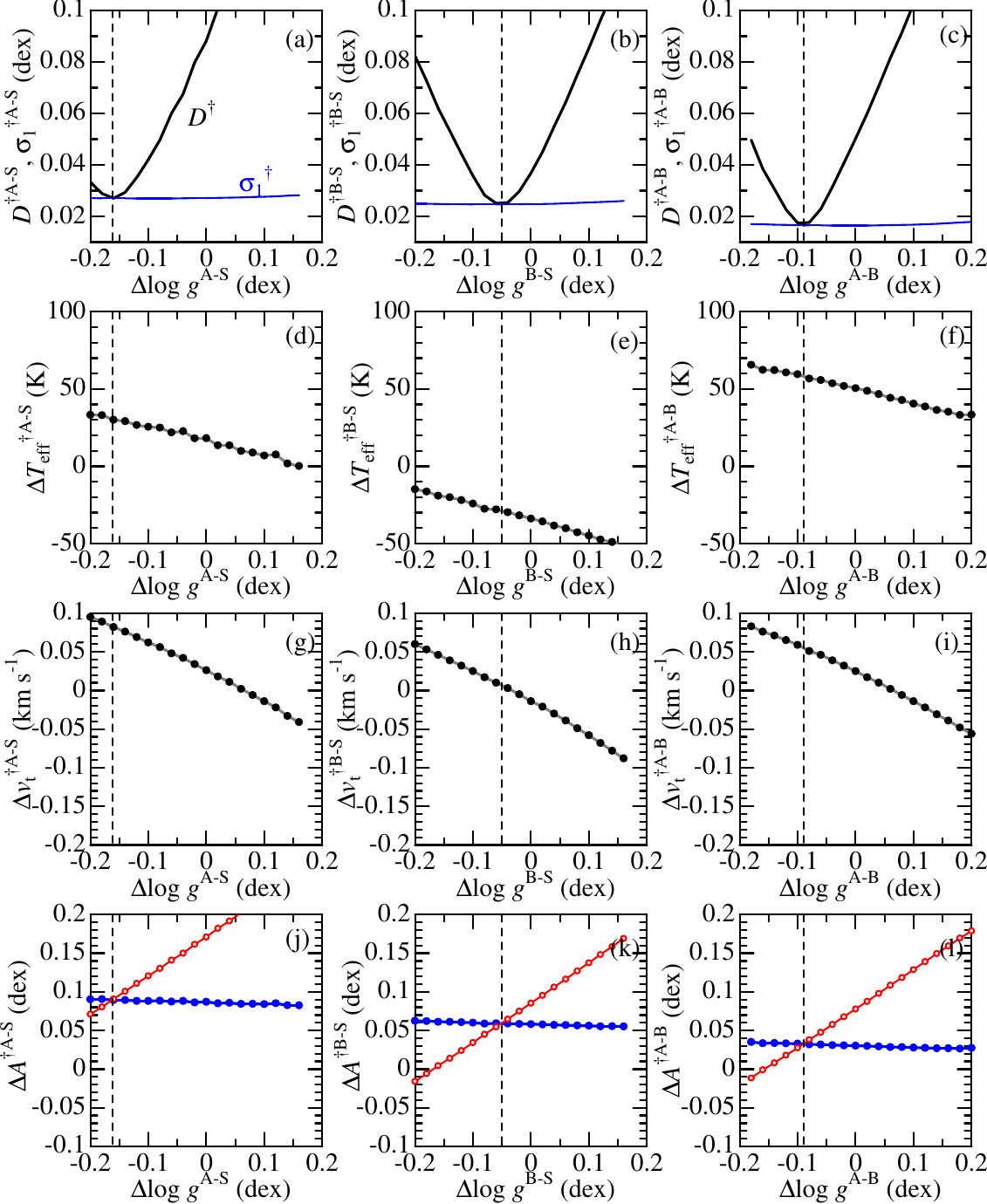}
\caption{
Top-row panels (a--c): $\sigma_{1}^{\dagger}(\Delta T_{\rm eff}^{\dagger},\Delta v_{\rm t}^{\dagger})$ 
and $D^{\dagger}(\Delta T_{\rm eff}^{\dagger},\Delta v_{\rm t}^{\dagger})$ plotted against 
$\Delta \log g$, where $(\Delta T_{\rm eff}^{\dagger},\Delta v_{\rm t}^{\dagger})$
is the (pseudo-)minimum position of $\sigma_{1}$ for each given $\Delta \log g$. 
2nd-row panels (d--f): Run of $\Delta T_{\rm eff}^{\dagger}$ with $\Delta \log g$.
3rd-row panels (g--i): Run of $\Delta v_{\rm t}^{\dagger}$ with $\Delta \log g$.
Bottom-row panels (j--l): Run of $\Delta A_{1}^{\dagger}$ (blue filled symbols) and 
$\Delta A_{2}^{\dagger}$ (red open symbols) with $\Delta \log g$.
In each panel, the position of final $\Delta \log g^{*}$ solution is indicated 
by the vertical dashed line. 
Otherwise, the same as in Fig.~2.
}
\label{fig3}
\end{minipage}
\end{figure}

As can be confirmed by comparing Fig.~2 and Fig.~3, the results derived from two approaches 
are consistent with each other. Fig.~4 demonstrates that the differential abundances 
($\Delta A_{i,1}$ and $\Delta A_{i,2}$) for each of the lines (corresponding to the established 
solutions of $\Delta T_{\rm eff}^{*}$, $\Delta \log g^{*}$, and $\Delta v_{\rm t}^{*}$)
reasonably satisfy the required three conditions described in Sect.~3.1.

\setcounter{figure}{3}
\begin{figure}
\begin{minipage}{80mm}
\includegraphics[width=8.0cm]{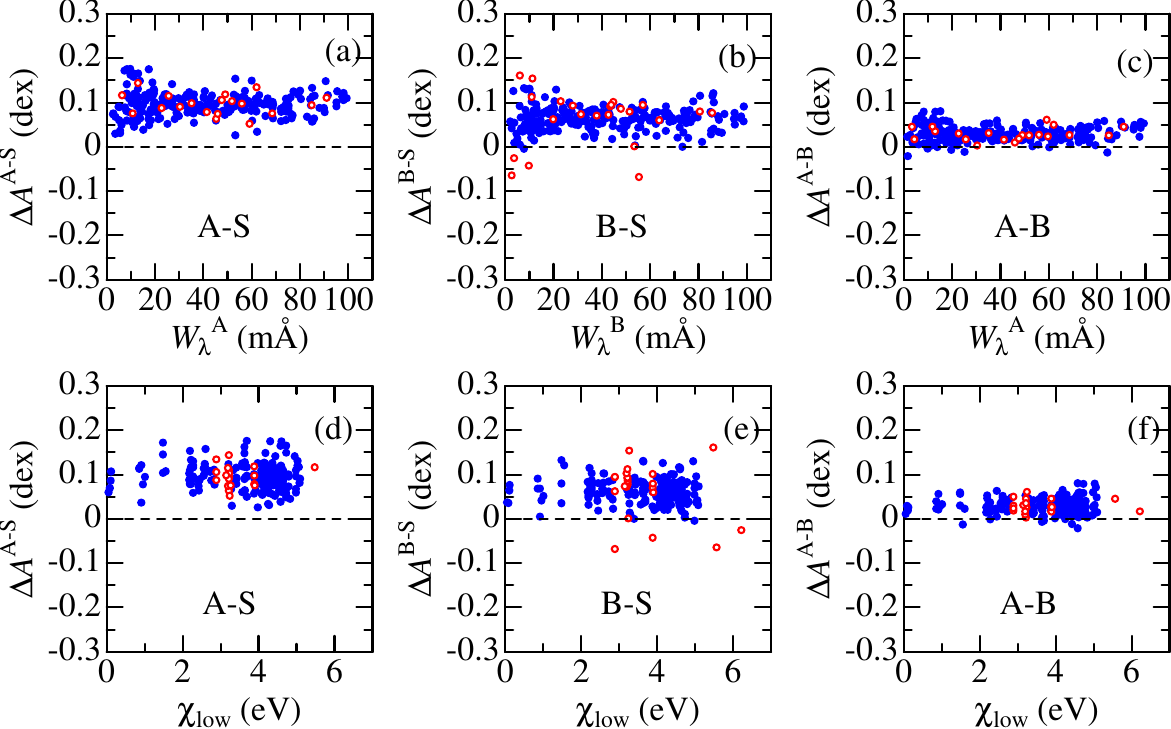}
\caption{
Upper panels (a--c): $\Delta A$ (differential Fe abundances between the target and 
the reference star corresponding to the final solutions of differential parameters) 
plotted against $W_{\lambda}$ (equivalent widths of Fe lines of the target star). 
Lower panels (d--f): $\Delta A$ plotted against $\chi_{\rm low}$ (lower excitation 
potential).  The blue filled and red open symbols correspond to Fe~{\sc i} and 
Fe~{\sc ii} lines, respectively. 
The left, centre, and right panels are for 16~Cyg~A $-$ Sun, 
16~Cyg~B $-$ Sun, and 16~Cyg~A $-$ 16~Cyg~B, respectively.
}
\label{fig4}
\end{minipage}
\end{figure}

\subsection{Error estimation}

The differential abundances ($\Delta A_{i,1}$/$\Delta A_{i,2}$) derived from 
the observed equivalent widths of $N_{1}/N_{2}$ lines of Fe~{\sc i}/Fe~{\sc ii} 
naturally contain random errors corresponding to the standard deviations 
($\sigma_{1}$/$\sigma_{2}$). How this abundance ambiguity affects the solutions 
of differential parameters was investigated as follows. (In the following, suffixes 
`1' and `2' are omitted for simplicity.) 

Randomly-generated noises of normal distribution ($e$) corresponding to $\sigma$ 
were added to the set of ``actual'' abundances to produce a new set of ``perturbed'' 
abundances as ($\Delta'A_{i} = \Delta A_{i} + e_{i}$, $i = 1, 2,\ldots, N$). 
Then, the same procedure as described in Sect.~3.2 is applied to this new set to 
obtain $\Delta' T_{\rm eff}$, $\Delta' \log g$, $\Delta' v_{\rm t}$, and $\Delta' A$.
This process was repeated 1000 times.

From these 1000 sets of perturbed solutions, standard deviations ($\sigma'_{T}$, 
$\sigma'_{g}$, $\sigma'_{v}$, and $\sigma'_{A}$) were calculated.
Then, the errors involved in the standard solutions 
($\Delta T_{\rm eff}^{*}$, $\Delta \log g^{*}$, $\Delta v_{\rm t}^{*}$, and 
$\Delta A^{*}$) determined in Sect.~3.2 may be regarded as ($\sigma'_{T}/\sqrt{2}$, 
$\sigma'_{g}/\sqrt{2}$, $\sigma'_{v}/\sqrt{2}$, and $\sigma'_{A}/\sqrt{2}$),
where the reason for dividing by $\sqrt{2}$ is to avoid duplication, since the original 
abundance set already has an intrinsic dispersion of $\sigma$. 

Such estimated solution errors are given in Table~2 (parenthesised values with $\pm$). 
The distributions of ($\Delta'_{n} T_{\rm eff}$, $\Delta'_{n} \log g$, 
$\Delta'_{n} v_{\rm t}$, and $\Delta'_{n} A, n =1, 2, \cdots, 1000$) are
also depicted in Fig.~5, where the error bars denote the values of $\pm \sigma'/\sqrt{2}$
mentioned above. 

\setcounter{figure}{4}
\begin{figure}
\begin{minipage}{80mm}
\includegraphics[width=8.0cm]{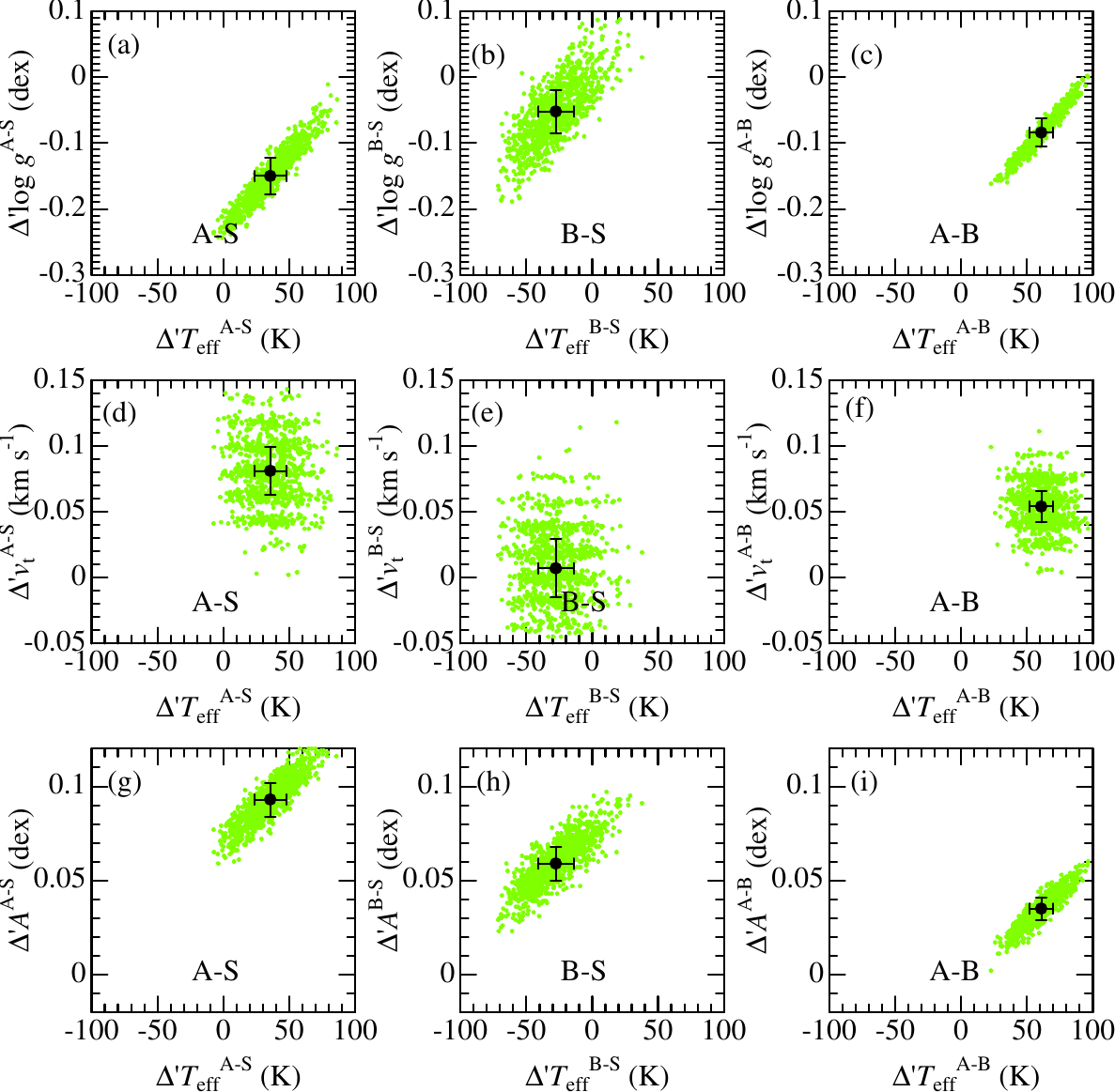}
\caption{Results of numerical experiments, where the differential 
analysis was repeated 1000 times by adding randomly-generated 
noises of normal distribution (corresponding to the actually 
measured standard deviation) to the set of standard Fe abundances.
In the top (a--c), middle (d--f), and bottom (g--i) panels are plotted 
the resulting $\Delta'_{n} \log g$, $\Delta'_{n} v_{\rm t}$, and 
$\Delta'_{n} A$ against $\Delta'_{n} T_{\rm eff}$ 
($n$ = 1, 2, $\cdots$, 1000). The mean of the distribution (practically 
the same as the standard solution) is indicated by a bullet
and the error bars indicate $\pm \sigma'/\sqrt{2}$ ($\sigma'$ is the
standard deviation of the distribution; see the main text for 
the reason of division by $\sqrt{2}$).
The left, centre, and right panels are for 16~Cyg~A $-$ Sun, 
16~Cyg~B $-$ Sun, and 16~Cyg~A $-$ 16~Cyg~B, respectively.
}
\label{fig5}
\end{minipage}
\end{figure}

\subsection{Mutual dependency of parameters and Fe abundances}

It is seen from Fig.~5 that close correlations exist between $\Delta T_{\rm eff}$, 
$\Delta \log g$, and $\Delta A$ (while $\Delta v_{\rm t}$ is rather independent unlike 
others); i.e., $\Delta \log g$ as well as $\Delta A$ tend to progressively increase
with an increase in $\Delta T_{\rm eff}$. Actually, these trends regarding
$\Delta T_{\rm eff}$, $\Delta \log g$, and $\Delta v_{\rm t}$ are also confirmed
from the characteristics in the contours of $D(\Delta T_{\rm eff}, \Delta \log g)$ and
$\sigma_{1}(\Delta T_{\rm eff}, \Delta v_{\rm t})$ (around the final solutions) 
shown in Fig.~6.

In order to understand the cause of these correlations, sensitivities of Fe abundances 
($A$) to changing each of the atmospheric parameters ($\partial A/\partial T_{\rm eff}$, 
$\partial A/\partial \log g$, $\partial A/\partial v_{\rm t}$, and 
$\partial A/\partial {\rm [Fe/H]}$) were estimated for the representative case of the Sun,
while perturbing each parameter (around the solar values of 5780~K, 4.44~dex, 1.0~km~s$^{-1}$,
and 0.0~dex) interchangeably by $\pm 50$~K, $\pm 0.1$~dex, $\pm 0.2$km~s$^{-1}$, 
and $\pm 0.1$~dex. The results are graphically displayed in Fig.~7, where $\partial A/\partial p$
($p$ is any of the four parameters) calculated for each line is plotted against
$W_{\lambda}$ and $\chi_{\rm low}$. The following trends can be read from this figure.
\begin{itemize}
\item
The typical values of  $\partial A/\partial T_{\rm eff}$ are $\sim +8 \times 10^{-4}$~dex/K 
(for Fe~{\sc i}) and $\sim -2 \times 10^{-4}$~dex/K (for Fe~{\sc ii}), which means that 
$A_{1}$ (Fe~{\sc i}) progressively increases while $A_{2}$ (Fe~{\sc ii}) decreases with 
$T_{\rm eff}$, and that $A_{2}$ is quantitatively much less sensitive to changing 
$T_{\rm eff}$ in comparison to $A_{2}$ (Figs.~7a and 7e). 
This trend is reasonably explained by the approximate expressions for the $T_{\rm eff}$-dependence 
of Fe abundance given by Eqs. (5) and (6) of Takeda et al. (2002): 
$A_{1} \simeq -(\chi^{\rm ion} - \chi_{\rm low})(5040/T_{\rm eff}) + {\rm const.}$ 
(only the essentially important exponential part is extracted here) 
and  
$A_{2} \simeq + \chi_{\rm low}(5040/T_{\rm eff})  + {\rm const.}$,
where $\chi^{\rm ion}$ is the ionisation potential of Fe~{\sc i} (7.87~eV) and
$\chi_{\rm low}$ is the lower excitation potential. That is, the difference 
between the typical values of $\chi^{\rm ion} - \chi_{\rm low}$ for Fe~{\sc i} 
($\sim $~4--5~eV on the average) and  $\chi_{\rm low}$ for Fe~{\sc ii} (mostly around 
$\sim 3$~eV) leads to a quantitative difference of $T_{\rm eff}$-sensitivity.   
\item
As to the $\log g$-sensitivity of $A$, $\partial A_{1}/\partial \log g$ tends to be 
very small (near to zero) while  $\partial A_{2}/\partial \log g$ is $\sim +0.4$~dex/dex,
though both deviating from this trend for stronger lines (cf. Fig.~7b).
Again, these characteristics can be understood by Eqs. (5) and (6) of Takeda et al. (2002),
which predict approximate relations that $A_{1}$ does not depend upon $\log g$ (because 
$g$-dependent electron density appearing in the line opacity is cancelled 
by that of the continuum H$^{-}$ opacity), while $A_{2} \sim +(1/3)\log g + {\rm const.}$ 
(due to $g^{1/3}$-dependence of the H$^{-}$ opacity).
\item
As expected, $\partial A/\partial v_{\rm t}$ is always negative and this 
$v_{\rm t}$-sensitivity progressively enhances with an increase in $W_{\lambda}$ (Fig.~7c). 
\item
Regarding the impact of model metallicity ([Fe/H] = [M/H]) on $A$,
$\partial A_{1}/\partial {\rm [Fe/H]}$ is almost zero and negligible 
(for the same reason as the case of $\log g$; i.e., the electron density 
effect tends to be cancelled out), while $A_{2}$ has some [Fe/H]-dependence as 
$\partial A_{2}/\partial {\rm [Fe/H]} \sim +0.1$~dex/dex. Still, this effect is quantitatively 
not so significant. For example, if [Fe/H] is changed by +0.1~dex, the increase in $A_{2}$  
is only $\sim +0.01$~dex. This is the reason why [Fe/H] was treated in an approximate
manner in Sect.~3.2 (see footnote~6).
\end{itemize}

\setcounter{figure}{5}
\begin{figure}
\begin{minipage}{80mm}
\includegraphics[width=8.0cm]{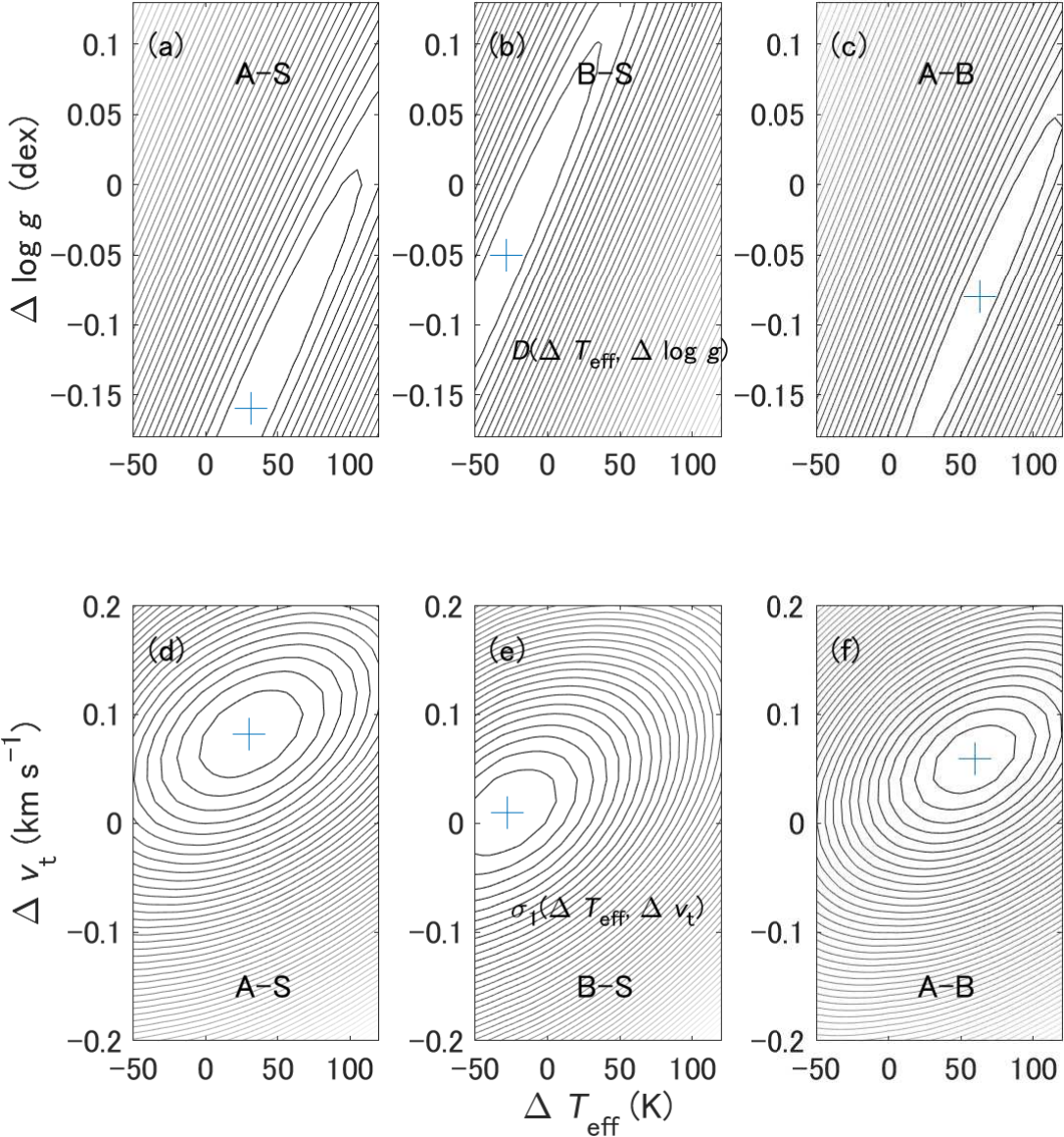}
\caption{Upper panels (a--c): graphical display of the contours of 
$D(\Delta T_{\rm eff}, \Delta\log g)$ at the final $\Delta v_{\rm t}^{*}$, 
where the position of $(\Delta T_{\rm eff}^{*}, \Delta\log g^{*})$ 
corresponding the minimum $D$ is indicated by a cross.
Lower panels (d--f): graphical display of the contours of 
$\sigma_{1}(\Delta T_{\rm eff}, \Delta v_{\rm t})$ at the final $\Delta\log g^{*}$,  
where the position of $(\Delta T_{\rm eff}^{*}, \Delta v_{\rm t}^{*})$ 
corresponding the minimum $D$ is indicated by a cross.
The left (a, d), centre (b, e), and right (c, f) panels are for 16~Cyg~A $-$ Sun, 
16~Cyg~B $-$ Sun, and 16~Cyg~A $-$ 16~Cyg~B, respectively.
}
\label{fig6}
\end{minipage}
\end{figure}

\setcounter{figure}{6}
\begin{figure}
\begin{minipage}{80mm}
\includegraphics[width=8.0cm]{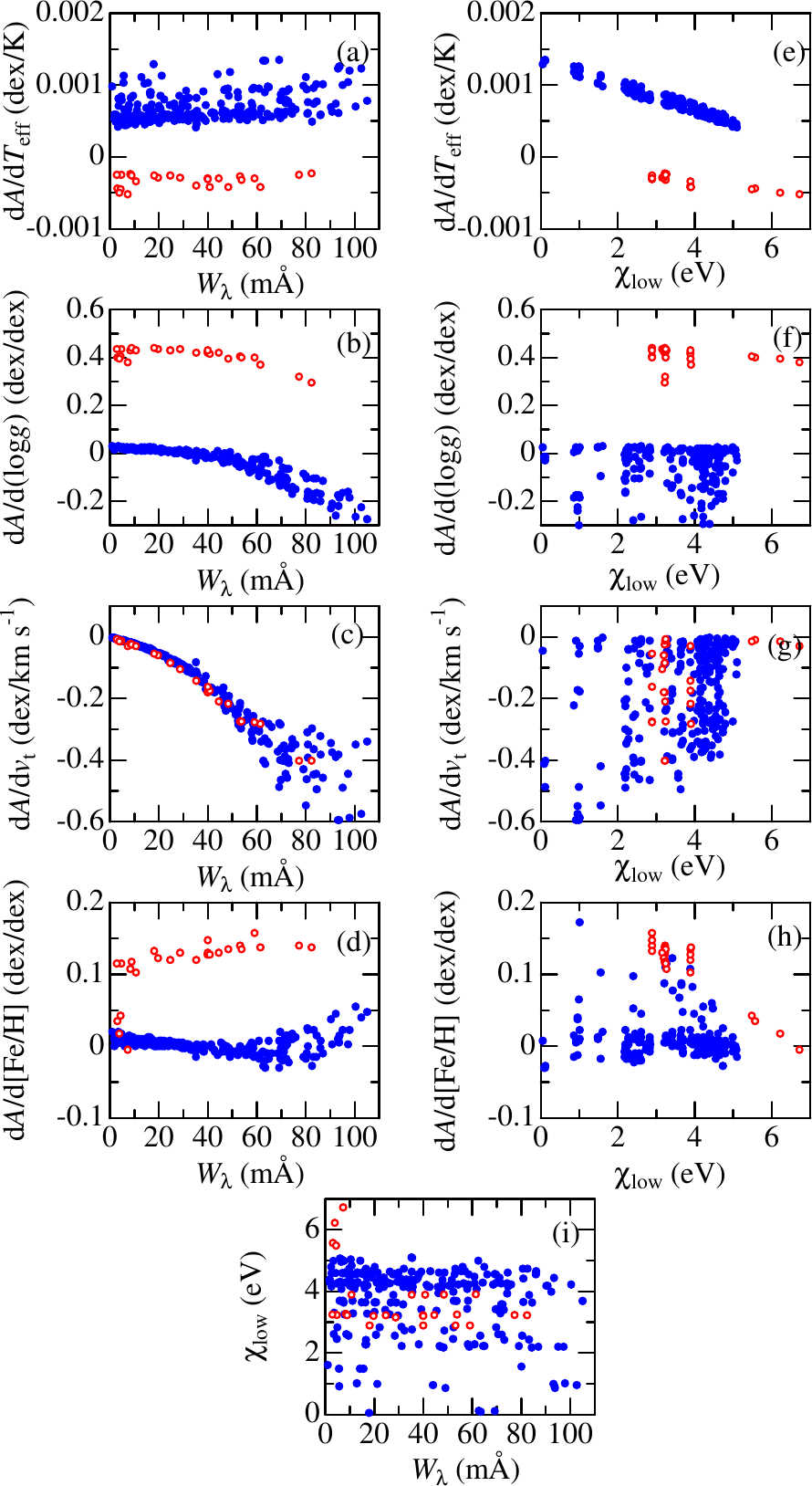}
\caption{Sensitivity of $A$ (Fe abundance) to changing the atmospheric parameters 
($T_{\rm eff}$, $v_{\rm t}$, $\log g$, and metallicity) calculated for the case of the Sun. 
The derivatives of $\partial A/\partial T_{\rm eff}$ (top row; a, e),
$\partial A/\partial \log g$ (2nd row; b, f),
$\partial A/\partial v_{\rm t}$ (3rd row; c, g), and
$\partial A/\partial [{\rm Fe/H}]$ (4th row; d, h),
 are plotted against $W_{\lambda}$ (left panels; a--d)
and $\chi_{\rm low}$ (right panels; e--h).
In the bottom panel (i) is shown the $W_{\lambda}$ vs. $\chi_{\rm low}$ distribution
of the solar Fe lines used in the present study.
 The blue filled and red open symbols correspond to Fe~{\sc i} and 
Fe~{\sc ii} lines, respectively.  
}
\label{fig7}
\end{minipage}
\end{figure}

Now, we can understand the correlations between $\Delta T_{\rm eff}$, $\Delta \log g$, 
and $\Delta A$ by combining the trends clarified above: (i) Variation of 
$\Delta A_{1}$ in response to changing $\Delta T_{\rm eff}$ is
$+8 \times 10^{-4} \delta (\Delta T_{\rm eff})$, 
while $\Delta A_{2}$ is rather $\Delta T_{\rm eff}$-independent (in a comparative sense).  
(ii) Conversely, $\Delta A_{1}$ is independent upon 
$\Delta \log g$, while $\Delta A_{2}$ is changed by $\sim +0.4 \delta (\Delta \log g)$
in response to perturbing $\Delta\log g$.
That is, since any change of $\Delta A_{1}$ is simply reflected also in $\Delta A_{2}$ 
due to the requirement of $\Delta A_{1} = \Delta A_{2} (= \Delta A)$,
combining (i) and (ii) leads to a relation  
\begin{equation} 
+8 \times 10^{-4} \delta (\Delta T_{\rm eff}) \sim 
+0.4 \delta (\Delta \log g),
\end{equation}
which describe the correlation between $\Delta T_{\rm eff}$ and $\Delta \log g$.
The sensitivity of $\Delta A$ to $\Delta T_{\rm eff}$ or $\Delta \log g$ is written as
\begin{equation}
\delta (\Delta A) \sim +8 \times 10^{-4} \delta (\Delta T_{\rm eff})
\end{equation}
and
\begin{equation}
\delta (\Delta A) \sim +0.4 \delta (\Delta \log g),
\end{equation}
where $\Delta T_{\rm eff}$ is in K and $\Delta \log g$ as well as $\Delta A$ are in dex. 
These relations reasonably explain the variations of $\delta (\Delta \log g) \sim +0.1$~dex
(Figs.~5a--5c) and $\delta (\Delta A) \sim +0.04$~dex (Figs.~5g--5i) for a change of 
$\delta (\Delta T_{\rm eff}) \sim +50$~K 

\section{Determination of elemental abundances}

Now that differential atmospheric parameters have been established for each of 
the star pairs (A--S, B--S, and A--B), we are ready to determine the abundance 
differences of various elements between the target and comparison stars,
where the conventional line-by-line differential analysis is applied to all lines
with measured equivalent widths (Sect.~2.2), though C, N, and O are separately 
treated by the spectrum synthesis analysis.
The parameters of the model atmospheres for the Sun, 16~Cyg~A, and 16~Cyg~B 
to be used for these analyses are summarised in Table~3 (which were chosen
based on the results given in Table~2). 

\setcounter{table}{2}
\begin{table}
\begin{minipage}{80mm}
\caption{Model parameters adopted for differential abundance determinations.}
\begin{center}
\begin{tabular}{ccccc}\hline
\hline
Star & $T_{\rm eff}$ & $\log g$ & $v_{\rm t}$ & [M/H]$^{*}$ \\
     & (K)  & (dex) & (km~s$^{-1}$) & (dex) \\
\hline
Sun       &      5780   &   4.44   &   1.00  &  ~0.00 \\       
16~Cyg~A  &      5811   &   4.28   &   1.08  &  +0.09 \\
16~Cyg~B  &     5751    &  4.39    &  1.01  &  +0.06 \\
\hline
\end{tabular}
\end{center}
$^{*}$Model metallicity represented by [Fe/H] $\equiv A({\rm Fe})-7.50$.
\end{minipage}
\end{table}

\subsection{Analysis of equivalent widths}

The equivalent widths of 531 lines measured for the Sun, 16~Cyg~A, and 16~Cyg~B 
(cf. Sect.~2.2), which are presented in the file ``ewlines.dat'', are classified
into  26 species: C~{\sc i} (4), O~{\sc i} (2), Na~{\sc i} (2), Al~{\sc i} (1),
Si~{\sc i} (20), Ca~{\sc i} (12), Sc~{\sc i} (2), Sc~{\sc ii} (9), Ti~{\sc i} (53),
Ti~{\sc ii} (6), V~{\sc i} (14), V~{\sc ii} (1), Cr~{\sc i} (26), Cr~{\sc ii} (5),
Mn~{\sc i} (3), Fe~{\sc i} (272), Fe~{\sc ii} (26), Co~{\sc i} (10), Ni~{\sc i} (51),
Cu~{\sc i} (1), Zn~{\sc i} (1), Y~{\sc ii} (4), La~{\sc ii} (1), Ce~{\sc ii} (3),
Pr~{\sc ii} (1), and Nd~{\sc ii} (1) (parenthesised are the number of lines).

The abundance $A_{i}$ was derived from $W_{i}$ of each line $i$ by using 
Kurucz's (1993) WIDTH9 program (though modified in various respects) for both 
the target and comparison stars to obtain the differential abundance
$\Delta A_{i} \equiv A_{i}({\rm target}) - A_{i}({\rm comparison})$.
Based on the resulting set of $(\Delta A_{{\rm X}i}, i=1, \cdots, N_{\rm X})$
for each species X, their mean ($\langle \Delta A_{\rm X}\rangle$), standard 
deviation ($\sigma_{\rm X}$), and mean error 
($\epsilon_{\rm X} \equiv \sigma_{\rm X}/\sqrt{N_{\rm X}}$) were computed,
where lines appreciably deviating from the main trend (judged by Chauvenet's 
criterion) were excluded from the mean. 
The complete results are presented in ``relabunds\_AtoS.dat'' (A--S),
``relabunds\_BtoS.dat'' (B--S), and ``relabunds\_AtoB.dat'' (A--B) of the 
supplementary material, where the adopted atomic parameters are also given.

\subsection{CNO abundances from spectrum fitting}

\subsubsection{Basic policy}

The abundances of C, N, and O (representative volatile elements particularly 
important in discussing the $T_{\rm c}$-dependence of the abundances) 
were determined from the line features of CH, NH, and OH molecules
in the blue--UV region by following the same procedure as adopted in Takeda (2023),\footnote{
In Takeda's (2023) investigation on the CNO abundances for 118 solar analogues, 
Sun (Vesta) and 16~Cyg~B were already included and analysed (but not 16~Cyg~A which 
was outside of the targets). It should be noted, however, that the model parameters
($T_{\rm eff}$, $\log g$, $v_{\rm t}$, and [M/H]) adopted there 
(5761~K, 4.43~dex, 1.00~km~s$^{-1}$, $-0.01$~dex for the Sun; 
5742~K, 4.32~dex, 1.01~km~s$^{-1}$, +0.08~dex for 16~Cyg~B)
were somewhat different from those used in this study (cf. Table~3).     
} which should be consulted for the details.

\subsubsection{Observed spectra}

The observational data used for this purpose are the high-dispersion spectra 
covering $\sim$~3000--4600~\AA\ (with a resolving power of $R \simeq 60000$) 
obtained in (UT) 2009 August 7 (16~Cyg~B), 2010 February 5 (Vesta), and 
2010 May 24 (16~Cyg~A) with the High Dispersion Spectrograph (HDS) of the 
Subaru Telescope (see Sect.~2 of Takeda et al. 2011 for more details).

\subsubsection{Fitting analysis}

As done in Takeda (2023), the spectrum fitting analysis was carried out 
in 12 regions in 4270--4330~\AA\ (CH), 11 regions in 3340--3390~\AA\ (NH), 
and 11 regions in 3100--3200~\AA\ (OH) (cf. Table~2 of Takeda 2023 for more details
and the accompanied online material in that paper for the adopted line data).  
However, since a considerable fraction of O abundances for 16~Cyg~A 
turned out to be unreliable and need to be rejected (cf. Sect.~4.2.4 below),
additional O abundance determinations were also implemented in newly adopted 
6 regions for the sake of compensation, as summarised in Table~4 
(the line data for these new 6 regions are given in ``lines\_OHregions.dat'' 
of the supplementary material).
How the observed and theoretically synthesised spectra could be fitted with 
each other in each region is displayed in Fig.~8 (12 regions for C), 
Fig.~9 (11 regions for N), Fig.~10 (11 regions for O), and Fig.~11 
(new 6 regions for O). 

\setcounter{table}{3}
\begin{table}
\begin{minipage}{80mm}
\begin{center}
\caption{New additional OH line regions for O abundance determination.}
\begin{tabular}{cccl}
\hline\hline
Region code & $\lambda_{1}$ & $\lambda_{2}$ & Varied abundances\\ 
\hline
OH3128 & 3127.76 & 3128.34 & Ti, OH                  \\
OH3133 & 3132.94 & 3133.68 & V, Fe, Zr, OH           \\
OH3167 & 3166.91 & 3167.53 & Fe, OH                  \\
OH3173 & 3172.68 & 3173.50 & Fe, OH                  \\
OH3241 & 3241.20 & 3242.46 & Ti, Fe, Y, OH           \\
OH3255 & 3255.00 & 3255.74 & Ti, Cr, Sc, OH          \\
\hline
\end{tabular}
\end{center}
These 6 OH line regions were newly included in this study
in addition to those adopted in Takeda (2023; cf. Table~2 therein).  
$\lambda_{1}$ and $\lambda_{2}$ in columns 2 and 3 are 
the starting and ending wavelengths (in \AA) of the spectral range
where the fitting analysis was done.
\end{minipage}
\end{table}

\setcounter{figure}{7}
\begin{figure}
\begin{minipage}{80mm}
\includegraphics[width=8.0cm]{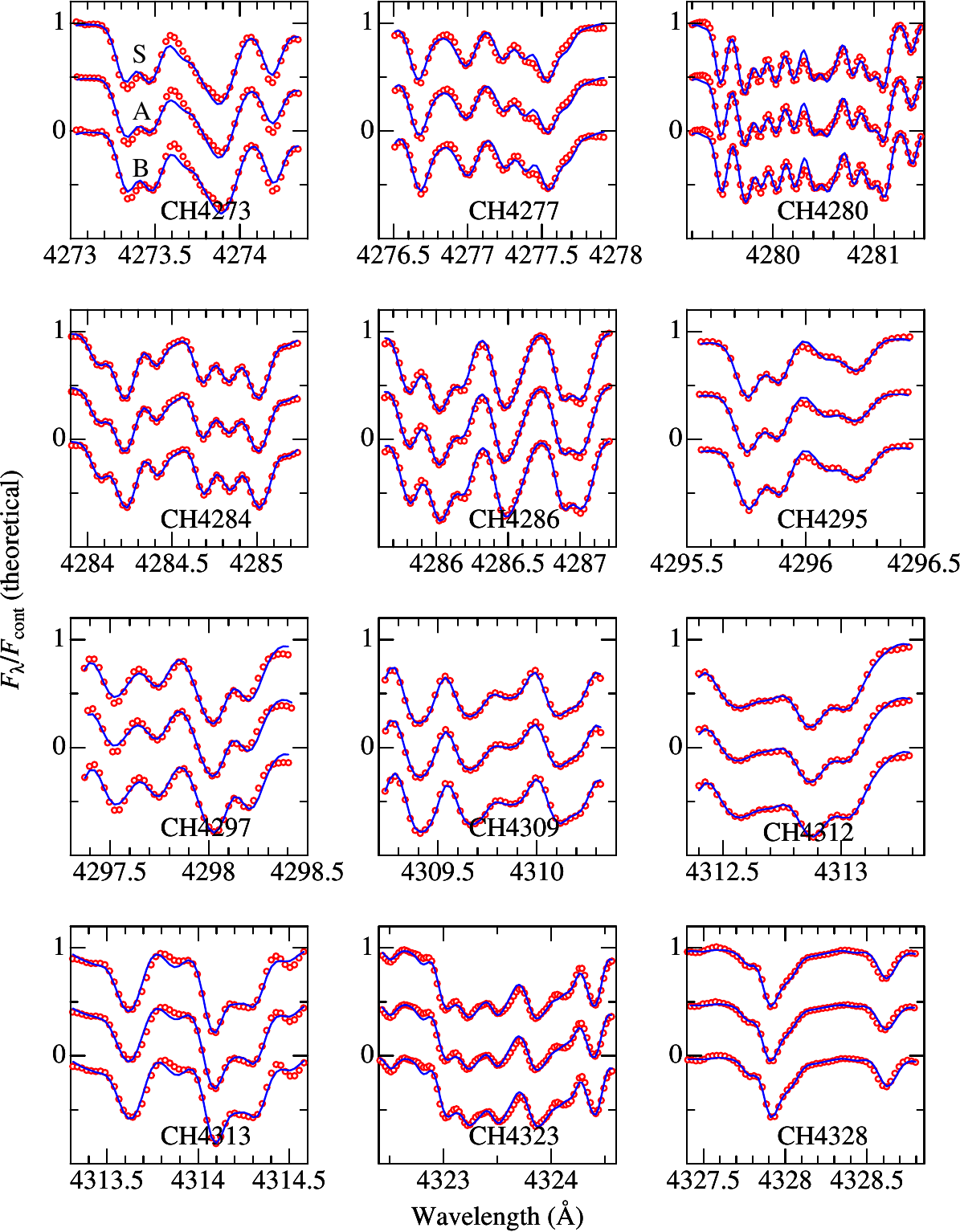}
\caption{
Observed and fitted theoretical spectra in each of the 12 regions 
(within 4270--4330~\AA), where C abundances were determined from CH lines. 
The corresponding region code (cf. Table~2 of Takeda 2023) is specified 
in each panel, where three spectra for the Sun (upper), 16~Cyg~A (middle)
and 16~Cyg~B (lower) are displayed side by side (each being vertically 
shifted by 0.5 relative to the adjacent one).
The observed and theoretical spectra are depicted by red open symbols 
and blue lines, respectively. The wavelength scale of the 
spectrum is adjusted to the laboratory frame, and the scale marked in 
the left ordinate corresponds to the theoretical residual flux 
($F^{\rm th}_{\lambda}/F^{\rm th}_{\rm cont}$) of the top (solar) spectra.
}
\label{fig8}
\end{minipage}
\end{figure}

\setcounter{figure}{8}
\begin{figure}
\begin{minipage}{80mm}
\includegraphics[width=8.0cm]{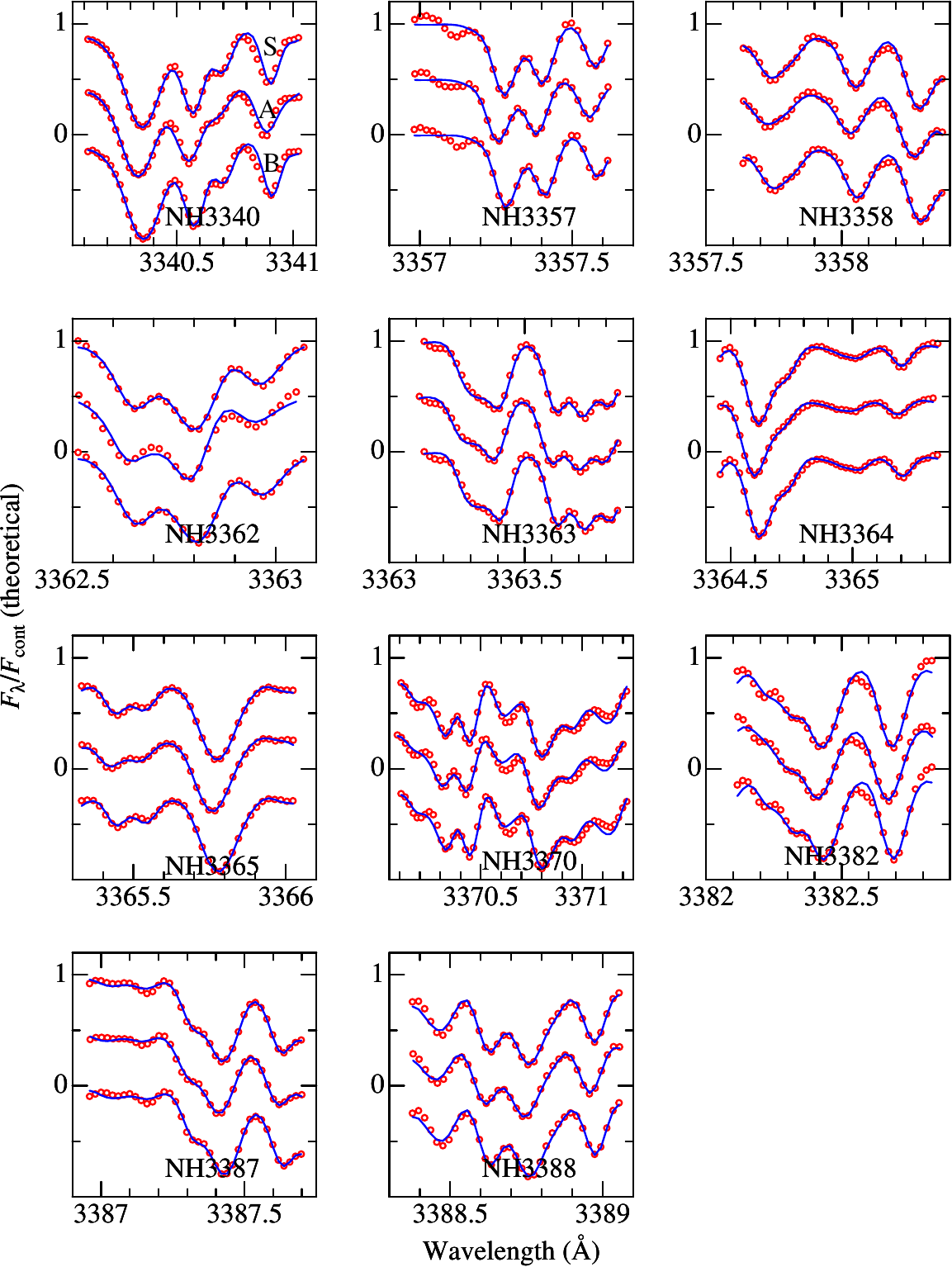}
\caption{
Observed and fitted theoretical spectra of the Sun, 16~Cyg~A, and 16~Cyg~B
in each of the 11 regions (within 3340--3390~\AA), where N abundances were determined 
from NH lines. Otherwise, the same as in Fig.~8.
}
\label{fig9}
\end{minipage}
\end{figure}

\setcounter{figure}{9}
\begin{figure}
\begin{minipage}{80mm}
\includegraphics[width=8.0cm]{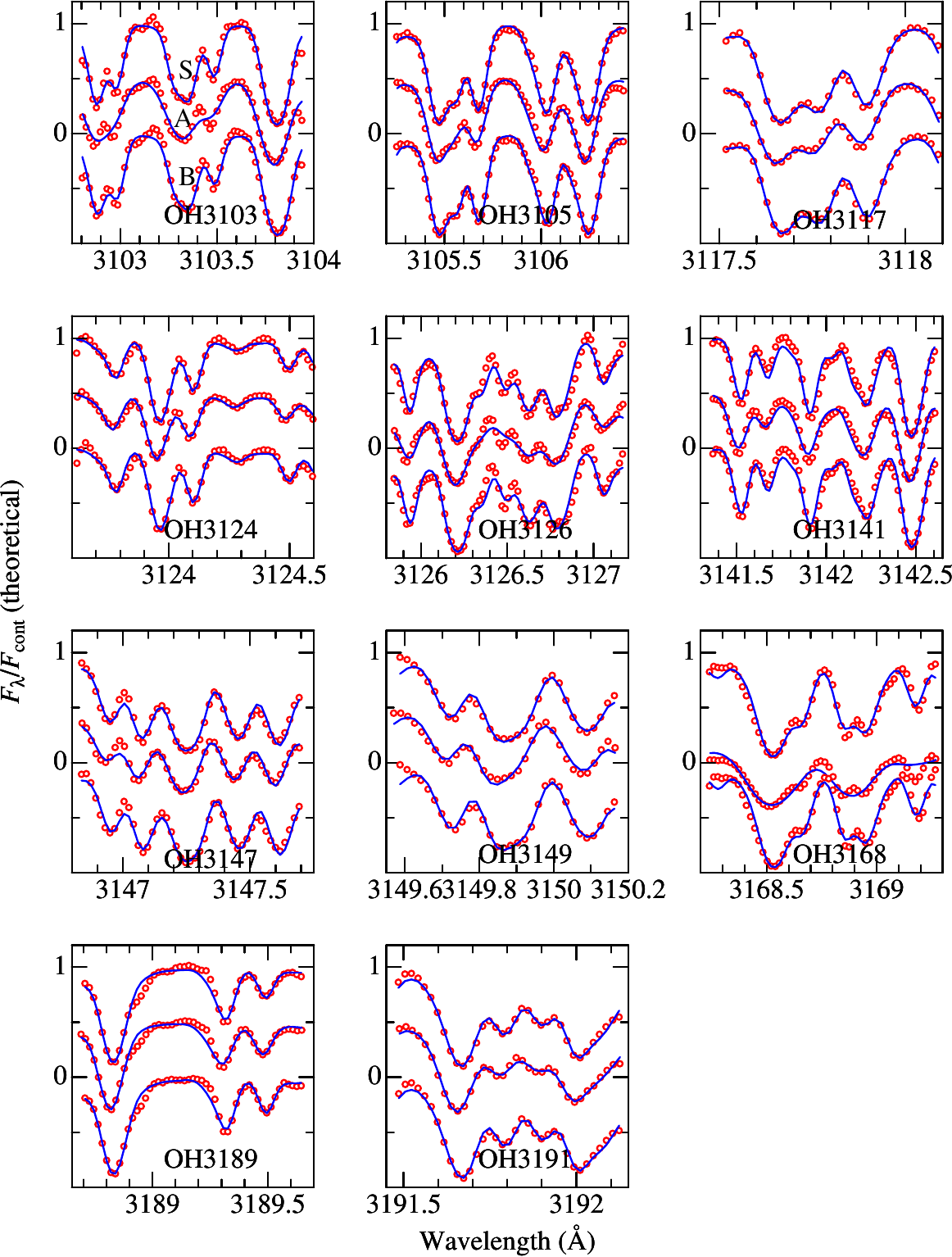}
\caption{
Observed and fitted theoretical spectra of the Sun, 16~Cyg~A, and 16~Cyg~B 
in each of the 11 regions (within 3100--3200~\AA), where O abundances were determined 
from OH lines. Otherwise, the same as in Fig.~8.
}
\label{fig10}
\end{minipage}
\end{figure}

\setcounter{figure}{10}
\begin{figure}
\begin{minipage}{80mm}
\includegraphics[width=8.0cm]{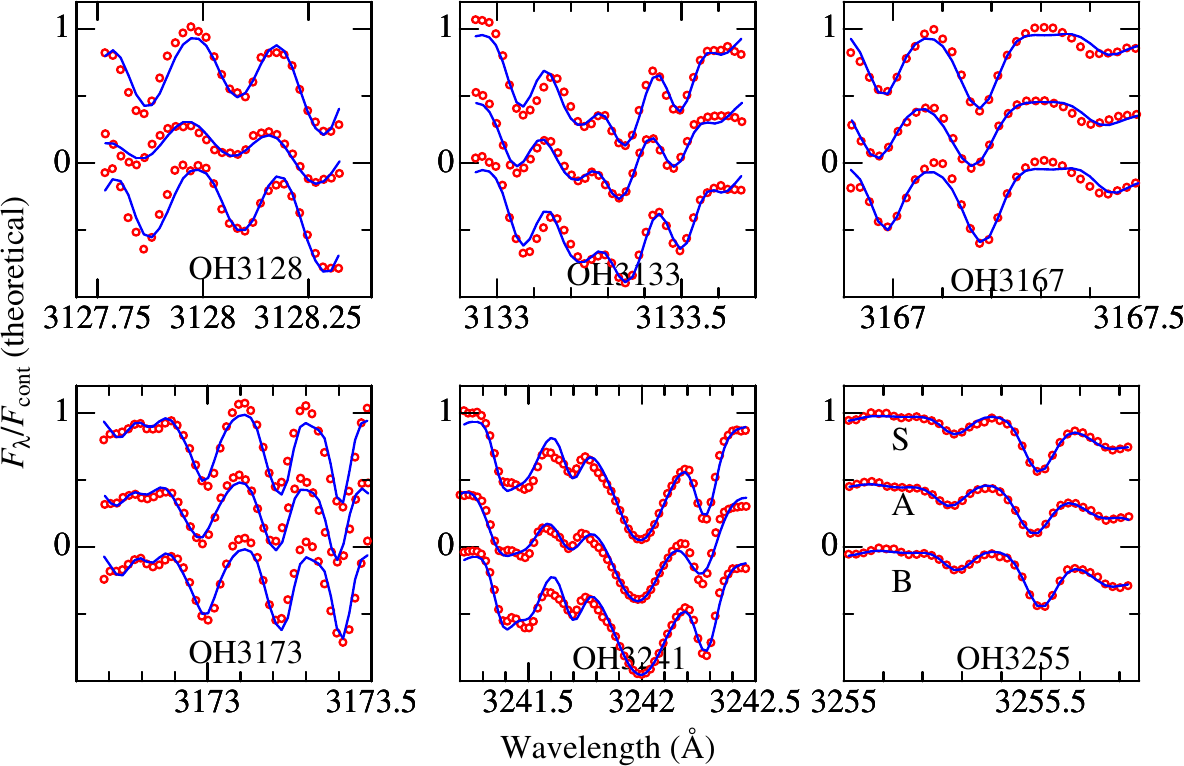}
\caption{
Observed and fitted theoretical spectra of the Sun, 16~Cyg~A, and 16~Cyg~B 
in the 6 OH line regions newly added in this study (cf. Table~4). 
Otherwise, the same as in Fig.~8.
}
\label{fig11}
\end{minipage}
\end{figure}

\subsubsection{Problematic O abundances for 16~Cyg~A}

Based on the resulting abundances of X (= C or N or O) at each region $k$ 
($A_{{\rm X}k}$), differential abundances $\Delta A_{{\rm X}k}$ were computed
for three star pairs (A--S, B--S, and A--B). The region-to-region comparisons of 
$\Delta A_{{\rm X}k}$ are displayed in Fig.~12.
It can be seen from Fig.~12 that $\Delta A_{\rm O}^{\rm A-S}$ 
and $\Delta A_{\rm O}^{\rm A-B}$ (i.e., differential oxygen abundances related to 
16~Cyg~A) show especially large dispersions from region to region (amounting to 
$\sim \pm 0.2$~dex), while reasonable consistency is observed for the other cases.
It may thus be concluded that these OH-based O abundances of 16~Cyg~A suffer 
considerable ambiguities and less reliable. 

Actually, quite a similar tendency is seen in the macrobroadening velocity $v_{\rm M}$ 
($e$-folding width of the Gaussian macrobroadening function $\propto \exp [-(v/v_{\rm M})^{2}]$, 
representing the combined effect of instrumental broadening, rotation, and macroturbulence), 
which is derived as a by-product of spectrum fitting. As illustrated in Fig.~13,
despite that these $v_{\rm M}$ solutions should in principle be region-independent 
(around $\sim 4$~km~s$^{-1}$), they show considerable scatter only in the  
OH region analyses of 16~Cyg~A (meaning that their solutions are not trustworthy). 

As such, some kind of problem must be involved in the observational data 
of 16~Cyg~A, which may be related with the fact that OH lines are in the shortest 
wavelength region ($\lambda \sim $~3100--3200~\AA; near to the detector sensitivity 
limit) being apt to suffer troubles (e.g., effect of stray light) because of low count 
levels. However, the reason why such a phenomenon is seen only in 16~Cyg~A (but 
not in the Sun and 16~Cyg~B) is not clear. 

Therefore, those O abundance solutions for 16~Cyg~A derived from OH regions 
yielding with anomalously large $v_{\rm M}$ ($> 5$~km~s$^{-1}$),
which are marked with red crosses in Fig.~12 and Fig.~13, were rejected from the outset 
without being included in the averaging.  

\setcounter{figure}{11}
\begin{figure}
\begin{minipage}{80mm}
\includegraphics[width=8.0cm]{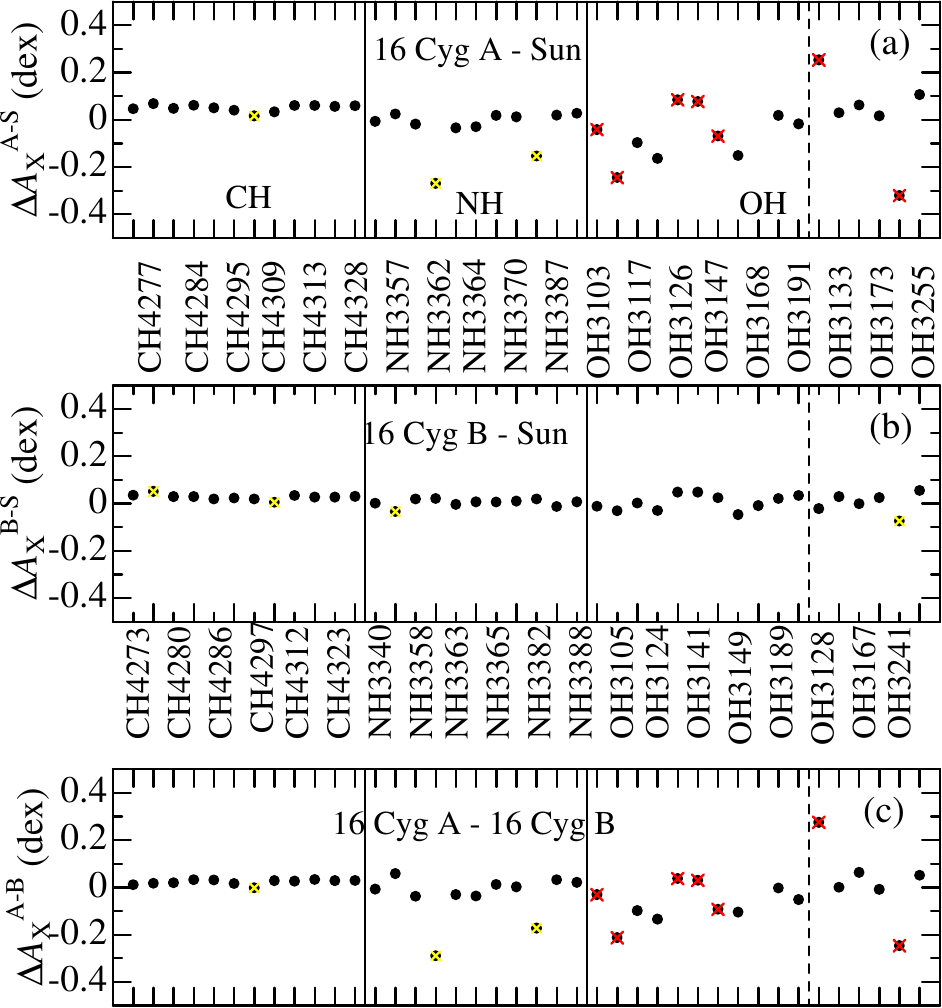}
\caption{
Comparison of the differential CNO abundances derived 
from each of the 40 regions. (a) $\Delta A_{\rm X}^{\rm A-S}$,
(b) $\Delta A_{\rm X}^{\rm B-S}$, and (c) $\Delta A_{\rm X}^{\rm A-B}$.
The red-crossed symbols (8 OH data related to 16~Cyg~A) are the pre-rejected data 
because of too large $v_{\rm M}$ (cf. Fig.~13), while those yellow-crossed  
ones correspond to the discarded data (judged to be outliers) based on 
Chauvenet's criterion. 
} 
\label{fig12}
\end{minipage}
\end{figure}

\setcounter{figure}{12}
\begin{figure}
\begin{minipage}{80mm}
\includegraphics[width=8.0cm]{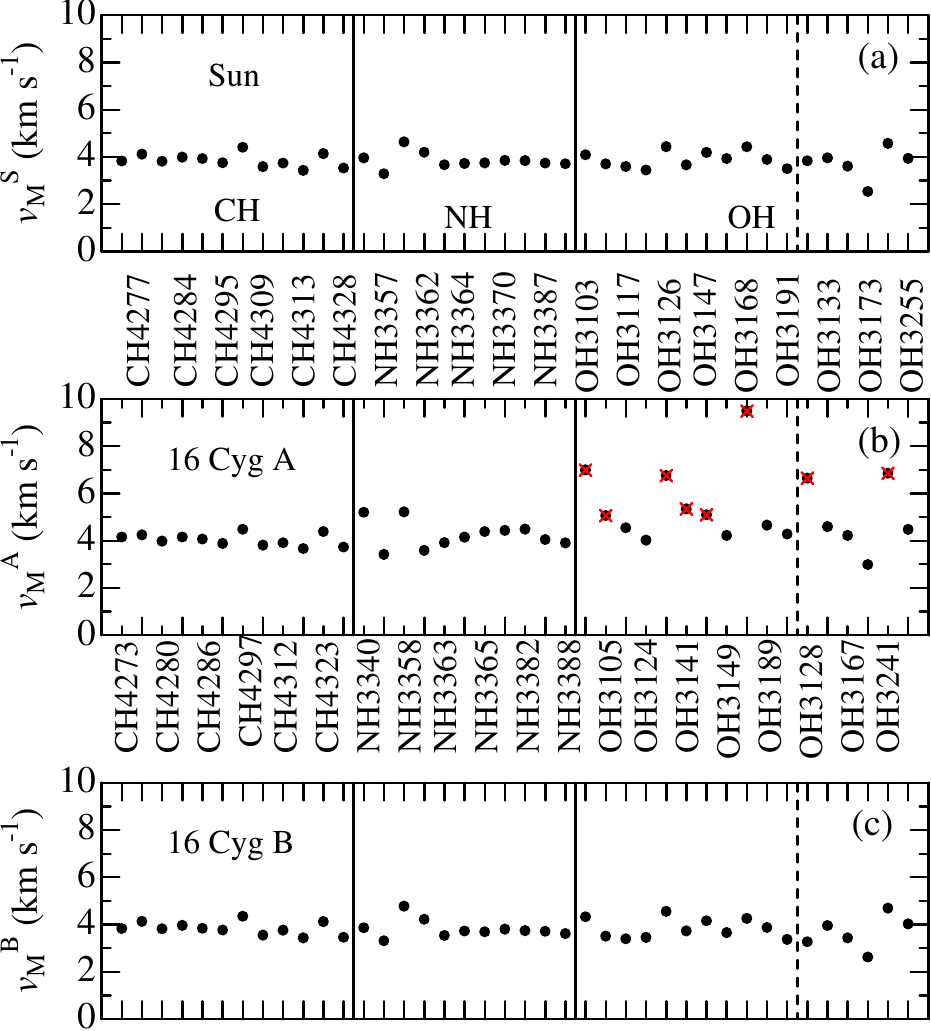}
\caption{
Comparison of $v_{\rm M}$ values ($e$-folding width of 
Gaussian macrobroadening function determined as a by-product of spectrum-fitting 
analysis) derived from each of the 40 regions (12, 11, and 11+6 for CH, NH, 
and OH). (a) Sun, (b) 16~Cyg~A, and (c) 16~Cyg~B. Red-crossed symbols are 
the rejected data (8 OH regions for 16~Cyg~A) because of anomalously large
$v_{\rm M}$ ($> 5$~km~s$^{-1}$).
} 
\label{fig13}
\end{minipage}
\end{figure}

\subsubsection{Final abundance results for C, N, and O} 

From the differential abundances $\Delta A_{{\rm X}k}$ for each region $k$
derived in Sect.~4.2.4 for three star pairs (A--S, B--S, and A--B; cf. Fig.~12),
their mean values were computed by averaging them over the regions, where
the outlier values judged by Chauvenet's criterion as well as the pre-rejected
8 abundances of O for 16~Cyg~A (yellow and red crosses in Fig.~12, respectively)
were discarded. The resulting mean differential abundances 
($\langle \Delta A_{\rm X} \rangle$), standard deviations ($\sigma_{\rm X}$), 
and mean errors ($\epsilon_{\rm X}$) are summarised in ``CNOabunds.dat'' of 
the supplementary online material.

An inspection of the results reveals that the mean oxygen abundances 
related to 16~Cyg~A ($\langle \Delta A_{\rm O}^{\rm A-S}\rangle$ 
and $\langle \Delta A_{\rm O}^{\rm A-B}\rangle$) rather deviate 
from those of C and N  (i.e., comparatively lower) and show an appreciably 
larger mean error ($\epsilon \sim$~0.02--0.03~dex) due to larger 
region-to-region abundance difference, as already mentioned in Sect.~4.2.4.
Therefore, the present OH-based results of 16~Cyg~A's lower O abundance 
by $\sim$~0.02--0.03~dex in comparison with the Sun as well as 16~Cyg~B 
should be viewed with caution.\footnote{
Unfortunately, the O abundances formally derived from the equivalent widths of 
forbidden [O~{\sc i}] 5577 and 6363 lines in Sect.~4.1 (A is more O-rich than B 
by 0.07~dex) should not be seriously taken because the blending effect by 
other lines (important in solar-type stars) was not taken into consideration.
More useful information may be obtained from the results of Takeda \& Honda (2005), 
who determined the oxygen abundances of 16~Cyg~A and B (relative to the Sun) from the 
equivalent widths of [O~{\sc i}] 6300, O~{\sc i} 6158, and O~{\sc i} 7773 lines, 
and obtained ([O/H]$^{\rm A}$, [O/H]$^{\rm B}$) = (+0.12,+0.20)$_{6300}$, 
(+0.06,+0.07)$_{6158}$, and (+0.07,+0.04)$_{7773}$.  
It should be noted, however, that $T_{\rm eff}$ values adopted by Takeda \& Honda (2005)
(5765~K/5795~K for A/B) are appreciably different from those in this study 
(5811~K/5751~K for A/B; cf. Table~3).
If the latter values are adopted for $T_{\rm eff}$, an inequality relation 
of [O/H]$^{\rm A}$ $<$ [O/H]$^{\rm B}$ is accomplished also for 
O~{\sc i} 6158/7773 lines, because abundances from these high-excitation lines are 
$T_{\rm eff}$-sensitive (typically $\sim -0.1$~dex for a change of +100~K). 
Then, the O-abundance differences between A and B would become almost consistent 
(at least qualitatively) between [O~{\sc i}], O~{\sc i}, and OH lines.} 
Therefore, in the discussion of $\Delta A$ vs. $T_{\rm c}$ trends to be presented
in Sect.~5.2 (i.e., determination of linear-regression lines in Sect.~5.2.1 or
comparison with the results of previous studies in Sect.~5.2.5), both of 
the O-included and O-excluded results are shown in parallel.

\section{Discussion}

\subsection{Metallicities of 16~Cyg~A and B}

As mentioned in Sect.~1, the primary motivation of this study was to examine whether 
any difference of metallicity (Fe abundance) exists between 16~Cyg~A (without planet) 
and 16~Cyg~B (hosting a giant planet), because recent investigations have reported 
a small but meaningful difference by a few hundredths dex (A$>$B) in contrast with 
Takeda's (2005) conclusion of almost the same metallicity (A$\simeq$B).
 
The differential analysis of stellar parameters carried out in Sect.~3 yielded  
$\Delta A_{\rm Fe}^{\rm A-B} = +0.036$ (Table~2).
Besides, the re-analysis of equivalent widths for all available Fe~{\sc i} lines 
(without restricting to lines of $W_{\lambda} < 100$~m\AA\ adopted in Sect.~3) done 
in Sect.~4.1 resulted in $\Delta A_{\rm Fe}^{\rm A-B} = +0.032$ (see Table~4).
Accordingly, it may be concluded that 16~Cyg~A is more metal-rich than 16~Cyg~B by
$+0.03$~dex, which is meaningful because the expected error is $\pm 0.01$~dex 
(including errors due to ambiguities in stellar parameters; cf. Sect.~5.2). 
This consequence is well consistent with the recent literature results 
published after 2010 (Schuler et al. 2011; Ram\'{\i}rez et al. 2011; 
Tucci Maia et al. 2014, 2019; Nissen et al. 2017; Ryabchikova et al. 2022; cf. Table~1).
Likewise, this is in reasonable agreement with the differential metallicities 
relative to the Sun ($\Delta A_{\rm Fe}^{\rm A-S} = +0.09$ and 
$\Delta A_{\rm Fe}^{\rm B-S} = +0.06$). 

I admit based on this result that my previous conclusion of no meaningful metallicity 
difference between A and B 
($\Delta A_{\rm Fe}^{\rm A-B} = 0.00 \pm 0.01$~dex; Takeda 2005) was incorrect.
Presumably, this is mainly due to the difference in $\Delta T_{\rm eff}$, which plays 
an important role in $\Delta A_{\rm Fe}$ as suggested from Table~1 (see also Fig.~3 of
Ryabchikova et al. 2022).
An inspection of Takeda's (2005) differential parameters between A and B reveals
that $\Delta T_{\rm eff}^{\rm A-B} = +39$~K (Table~1) is smaller
than the present result of +63~K (Table~2). This difference of $-24$~K correspond to 
an underestimation of $-0.02$~dex in $\Delta A_{\rm Fe}$ according to Eq.~(7), which would have 
erroneously shifted the actual $\Delta A_{\rm Fe}$ (+0.03~dex) downwards near to zero.

Then, why did Takeda's (2005) analysis yielded such a result? 
As a test, I reanalysed those old $W_{\lambda}$ data by using the same manner
as in Sect.~3, and obtained $\Delta T_{\rm eff}^{\rm A-B} = +50.6$~K, 
$\Delta \log g^{\rm A-B} = -0.044$~dex, 
$\Delta v_{\rm t}^{\rm A-B} = +0.098$~km~s$^{-1}$, 
and $\Delta A_{\rm Fe}^{\rm A-B} = +0.005$~dex.
That is, almost same metallicity for both A and B was reproduced as in Takeda (2005) 
(even though the solutions for $\Delta T_{\rm eff}^{\rm A-B}$ and 
$\Delta\log g^{\rm A-B}$ show small differences). This suggests that 
the observational $W_{\lambda}$ data used in Takeda (2005) were not adequate. 
I suspect that inclusion of Fe lines in the longer wavelength region (up to $\lesssim 7200$~\AA)
in my previous analysis might have been responsible for this problem,
where the spectra are of comparatively lower quality (due to insufficient S/N ratio,
or contamination by telluric lines).

\subsection{$T_{\rm c}$-dependence of elemental abundances}

\subsubsection{$\Delta A_{\rm X}$ vs. $T_{\rm c}$ relations}

In Sect.~4, the differential abundances ($\Delta A_{\rm X}$) for three star pairs 
(A--S, B--S, and A--B) were derived from equivalent widths for various elements
(26 species of 21 elements; cf. Sect.~4.1) and those of CNO were determined 
by applying the synthetic spectrum fitting method to blue--UV regions comprising 
lines of CH, NH, and OH molecules (Sect.~4.2).  

Let us examine how these $\Delta A_{\rm X}$ depend upon 
$T_{\rm c}$ (condensation temperature). Here, we restrict ourselves to
only those species for which 10 or more lines/regions are available, 
in order to secure the reliability of the resulting mean abundances as much as possible. 
That is, hydride molecules of volatile elements (CH, NH, OH) and neutral species of 
refractory elements (Si~{\sc i}, Ca~{\sc i}, Ti~{\sc i}, V~{\sc i}, Cr~{\sc i}, 
Fe~{\sc i}, Co~{\sc i}, and Ni~{\sc i}) (cf. Table~5), which are also 
advantageous because their abundance sensitivities to $T_{\rm eff}$ or $\log g$ 
are rather similar to each other (cf. Table~6) and thus their ``mutual'' 
relationships are not much influenced by errors in these parameters.

The mean differential abundances averaged over the available lines/regions 
($\Delta A_{\rm X}$) along with the mean errors ($\epsilon_{\rm X}$) 
for these 10 elements (X) are summarised in Table~5. Further, these 
resulting $\Delta A_{\rm X}$ values are plotted against $T_{\rm c}$ 
in Figs.~14a (A--S), 14b (B--S), and 14c (A--B), where the linear-regression
relations between $\Delta A_{\rm X}$ (in dex) and $T_{\rm c}$ (in K) 
determined by the least-squares analysis are also depicted: 
$\Delta A_{\rm X}^{\rm A-S} = +5.75\times 10^{-3} + 7.09\times 10^{-5} T_{\rm c}$,
$\Delta A_{\rm X}^{\rm B-S} = +1.11\times 10^{-2} + 4.51\times 10^{-5} T_{\rm c}$, and
$\Delta A_{\rm X}^{\rm A-B} = -3.86\times 10^{-3} + 2.37\times 10^{-5} T_{\rm c}$.
If oxygen abundances ($\Delta A_{\rm O}$) are excluded because of their
problematic nature (cf. Sect.~4.2), these relations are somewhat changed as follows:
$\Delta A_{\rm X(no O)}^{\rm A-S} = +2.52\times 10^{-2} + 5.73\times 10^{-5} T_{\rm c}$,
$\Delta A_{\rm X(no O)}^{\rm B-S} = +1.64\times 10^{-2} + 4.13\times 10^{-5} T_{\rm c}$, and
$\Delta A_{\rm X(no O)}^{\rm A-B} = +1.17\times 10^{-2} + 1.28\times 10^{-5} T_{\rm c}$.

\setcounter{table}{4}
\begin{table*}
\begin{minipage}{150mm}
\caption{Elemental abundance differences between the Sun, 16~Cyg~A, and 16~Cyg~B.}
\begin{center}
\begin{tabular}{ccccccccccc}\hline
\hline
$Z$ & Elem. & Line & $T_{\rm c}$ & $n_{\rm t}$ & $\Delta A^{\rm A-S}$ & 
$\epsilon^{\rm A-S}$ & $\Delta A^{\rm B-S}$ & $\epsilon^{\rm B-S}$ & 
$\Delta A^{\rm A-B}$ & $\epsilon^{\rm A-B}$ \\
(1) & (2) & (3) & (4) & (5) & (6) & (7) & (8) & (9) & (10) & (11)\\
\hline
 6 & C  &  CH    &    40 &   12 &   +0.053  &  0.003 &   +0.027 &   0.002 &   +0.025  &  0.002 \\
 7 & N  &  NH    &   123 &   11 &   +0.001  &  0.008 &   +0.007 &   0.003 &   +0.001  &  0.010 \\
 8 & O  &  OH    &   180 &   17 &  $-0.022$ &  0.030 &   +0.008 &   0.008 &  $-0.032$ &  0.022 \\
14 & Si &  Si~{\sc i}  &  1310 &   20 &   +0.128  &  0.005 &   +0.091 &   0.005 &   +0.029  &  0.005 \\
20 & Ca &  Ca~{\sc i}  &  1517 &   12 &   +0.112  &  0.004 &   +0.063 &   0.004 &   +0.044  &  0.003 \\
22 & Ti &  Ti~{\sc i}  &  1582 &   53 &   +0.105  &  0.005 &   +0.074 &   0.005 &   +0.032  &  0.005 \\
23 & V  &  V~{\sc i}   &  1429 &   14 &   +0.085  &  0.017 &   +0.064 &   0.009 &   +0.021  &  0.007 \\
24 & Cr &  Cr~{\sc i}  &  1296 &   26 &   +0.094  &  0.007 &   +0.073 &   0.005 &   +0.022  &  0.008 \\
26 & Fe &  Fe~{\sc i}  &  1334 &  272 &   +0.092  &  0.002 &   +0.059 &   0.002 &   +0.032  &  0.001 \\
27 & Co &  Co~{\sc i}  &  1352 &   10 &   +0.131  &  0.025 &   +0.097 &   0.021 &   +0.032  &  0.004 \\
28 & Ni &  Ni~{\sc i}  &  1353 &   51 &   +0.101  &  0.003 &   +0.078 &   0.003 &   +0.024  &  0.003 \\
\hline
\end{tabular}
\end{center}
(1) Atomic number. (2) Element species. (3) Adopted lines. (4)  Condensation temperature (in K) 
taken from Table~8 (50\% $T_{\rm c}$) in Lodders (2003).  (5) Number of available lines 
(or spectral regions for the case of CNO). (6) Averaged abundance difference 
(16~Cyg~A $-$ Sun; dex). (7) Mean error (dex). (8) Averaged abundance difference 
(16~Cyg~B $-$ Sun; dex)  (9) Mean error (dex). (10) Averaged abundance difference 
(16~Cyg~A $-$ 16~Cyg~B; dex). (11) Mean error (dex).
\end{minipage} 
\end{table*}

\subsubsection{Expected errors in $\Delta A_{\rm X}$}

Although the error bars attached to each of the symbols in these figures 
are simply $\pm \epsilon_{\rm X}$ (mean error), errors due to uncertainties in atmospheric 
parameters should also be added, which are typically $\sim \pm 0.01$~dex
for most of the elements (cf. $|\langle\delta_{Tgv}\rangle|$ values given in Table~6,
which are abundance errors corresponding to typical parameter uncertainties of
10~K, 0.02~dex, and 0.02~km~s$^{-1}$).
Accordingly, combined errors may be expressed as the root-sum-square 
of these two, or roughly approximated as $\sim$~max(0.01, $\epsilon_{\rm X}$). 
This means that actual errors involved in $\Delta A_{\rm X}$ values are 
$\simeq \pm 0.01$~dex in most cases, though more enhanced values of 
$\simeq \pm \epsilon_{\rm X}$ may be relevant for several cases of particularly 
large $\epsilon_{\rm X}$ (i.e., 
$\epsilon_{\rm O}^{\rm A-S} \simeq 0.03$,
$\epsilon_{\rm O}^{\rm A-B} \simeq 0.02$,
$\epsilon_{\rm V}^{\rm A-S} \simeq 0.02$,
$\epsilon_{\rm Co}^{\rm A-S} \simeq 0.03$, and
$\epsilon_{\rm Co}^{\rm B-S} \simeq 0.02$).

\setcounter{table}{5}
\begin{table}
\begin{minipage}{80mm}
\caption{Sensitivity of abundances to changing atmospheric parameters 
(evaluated for the case of the Sun).}
\begin{center}
\begin{tabular}{cccccc}\hline
\hline
Elem. & Line & $\langle\delta_{T+}\rangle$ & $\langle\delta_{g+}\rangle$ & 
$\langle\delta_{v+}\rangle$ & $|\langle\delta_{Tgv}\rangle|$ \\
(1) & (2) & (3) & (4) & (5) & (6) \\  
\hline
C  &  CH         &  +0.0073 & $-0.0022$ & $-0.0001$ &  0.008\\
N  &  NH         &  +0.0092 & $-0.0041$ & $+0.0001$ &  0.010\\
O  &  OH         &  +0.0102 & $-0.0042$ & $-0.0003$ &  0.011\\
Si &  Si~{\sc i} &  +0.0016 & $+0.0017$ & $-0.0006$ &  0.002\\
Ca &  Ca~{\sc i} &  +0.0072 & $-0.0048$ & $-0.0044$ &  0.010\\
Ti &  Ti~{\sc i} &  +0.0100 & $-0.0012$ & $-0.0023$ &  0.010\\
V  &  V~{\sc i}  &  +0.0109 & $-0.0004$ & $-0.0004$ &  0.011\\
Cr &  Cr~{\sc i} &  +0.0077 & $-0.0018$ & $-0.0030$ &  0.008\\
Fe &  Fe~{\sc i} &  +0.0070 & $-0.0016$ & $-0.0036$ &  0.008\\
Co &  Co~{\sc i} &  +0.0064 & $+0.0013$ & $-0.0009$ &  0.007\\
Ni &  Ni~{\sc i} &  +0.0056 & $+0.0010$ & $-0.0034$ &  0.007\\
\hline
\end{tabular}
\end{center}
(1) Element species. (2) Adopted lines. (3) Averaged abundance changes (dex) 
in response to $T_{\rm eff}$ variation of +10~K. (4) Averaged abundance changes (dex) 
in response to $\log g$ variation of +0.02~dex. (5) Averaged abundance changes (dex) 
in response to $v_{\rm t}$ variation of +0.02~km~s$^{-1}$.
(6) Root-sum-square of three $\langle\delta\rangle$ values 
$\bigl(\equiv \sqrt{\langle\delta_{T+}\rangle^{2}+\langle\delta_{g+}\rangle^{2}+
 \langle\delta_{v+}\rangle^{2}}\bigr)$. Although this way of combining three 
$\langle\delta\rangle$'s is not strictly reasonable (as uncertainties 
in $T_{\rm eff}$ and $\log g$ are not independent; see Sect.~3.4), it is 
sufficient because the effect of $T_{\rm eff}$ is dominant in any event 
($|\langle\delta_{Tgv}\rangle| \sim |\langle\delta_{T+}\rangle|$)   
\end{minipage}
\end{table}

\subsubsection{$\Delta A_{\rm X}^{\rm A-S}$ and $\Delta A_{\rm X}^{\rm B-S}$}

As to the differential abundances relative to the Sun,
both $\Delta A_{\rm X}^{\rm A-S}$ and 
$\Delta A_{\rm X}^{\rm B-S}$ 
tend to systematically increase with $T_{\rm c}$ in global perspective,
because $\Delta_{\rm X}$'s of refractory elements ($T_{\rm c} \sim $~1300--1600~K)
are $\lesssim +0.1$ while those of volatile CNO ($T_{\rm c} \sim $~40--180~K) 
distribute around $\sim 0.0$ (Figs.~14a and 14b). 

This tendency may be regarded as mainly due to unusual abundance 
characteristics of the Sun (reference star), because solar abundances 
are rather atypical (as revealed by high-precision differential analysis 
of solar twins relative to the Sun; cf. Mel\'{e}ndez et al. 2009) 
in the sense that refractory (high $T_{\rm c}$) elements are comparatively 
deficient relative to those of volatile (low $T_{\rm c}$) ones by $\lesssim 0.1$~dex.

However, the local behaviours of CNO abundances are somewhat complex 
because $\Delta A_{\rm C}$ is larger than $\Delta A_{\rm N}$ or $\Delta A_{\rm O}$ 
(i.e., tending to decrease with an increase in $T_{\rm c}$), and this trend is more 
conspicuous in 16~Cyg~A (Fig.~14a) than in 16~Cyg~B (Fig.~14b).  

\setcounter{figure}{13}
\begin{figure}
\begin{minipage}{80mm}
\includegraphics[width=8.0cm]{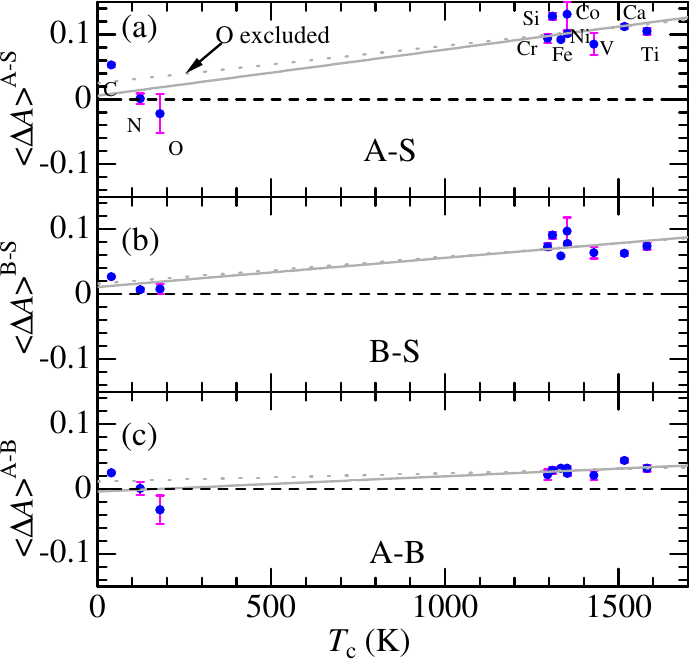}
\caption{
The mean differential abundances ($\Delta A_{\rm X}$) for 11 
elements (C, N, O, Si, Ca, Ti, V, Cr, Fe, Co, and Ni) presented in Table~5 
are plotted against $T_{\rm c}$ (condensation temperature).
The error bar attached to each symbol indicate $\pm\epsilon_{\rm X}$ (mean error) 
given in Table~5. (In addition, errors due to parameter uncertainties 
should also be considered, as described in Sect.~5.2.2.)
The linear-regression lines determined by least-squares analysis
are also drawn by gray solid line (for all 11 elements) and dotted line 
(for 10 elements excluding O) in each panel.
Panels (a), (b), and (c) are for 16~Cyg~A $-$ Sun ($\Delta A_{\rm X}^{\rm A-S}$), 
16~Cyg~B $-$ Sun ($\Delta A_{\rm X}^{\rm B-S}$), and 16~Cyg~A $-$ 16~Cyg~B
($\Delta A_{\rm X}^{\rm A-B}$), respectively.
} 
\label{fig14}
\end{minipage}
\end{figure}
    
\subsubsection{$\Delta A_{\rm X}^{\rm A-B}$}

Although the $T_{\rm c}$-dependence behaviours of $\Delta A_{\rm X}^{\rm A-S}$ 
and $\Delta A_{\rm X}^{\rm B-S}$ are qualitatively similar as mentioned above,
some difference does exist between A and B, since the abundances of refractory 
elements for A are slightly higher than B by $\sim$~+0.02--0.04~dex 
while those of volatile CNO do not show such a trend (scattered around $\sim 0$ 
for both A and B), 
which results in a positive gradient of $+2.4 \times 10^{-5}$~dex/K
(or $+1.3 \times 10^{-5}$~dex/K if O is excluded)  
in the $\Delta A_{\rm X}^{\rm A-B}$ vs. $T_{\rm c}$ relation (Fig.~14c).
This positive slope is observed not only for the global trend of 
volatile+refractory elements ($0 \lesssim T_{\rm c} \lesssim 1600$~K) but also 
for the local trend of refractory elements ($1300 \lesssim T_{\rm c} \lesssim 1600$~K).

\subsubsection{Comparison with previous studies}

The systematically increasing trend of $\Delta A_{\rm X}^{\rm A-S}$ 
and $\Delta A_{\rm X}^{\rm B-S}$ with $T_{\rm c}$ (Sect.~5.2.3) is 
more or less consistent with the results of other related studies done 
after 2010 (cf. Table~1). Regarding the behaviour of differential abundances 
between A and B ($\Delta A_{\rm X}^{\rm A-B}$), although any clear 
trend was not reported in the earlier work (Schuler et al. 2011; 
Ram\'{\i}rez et al. 2011), a tendency of increasing 
$\Delta A_{\rm X}^{\rm A-B}$ with $T_{\rm c}$ (similar to that  
obtained in Sect.~5.2.4) was concluded in the more recent studies over the past 
decade (Tucci Maia et al. 2014, 2019; Nissen et al. 2017; Ryabchikova et al. 2022).
Accordingly, the consequence of the present investigation is almost a reconfirmation 
of what has been reported by these authors in the qualitative sense. 

However, from a quantitative point of view, the gradient of the $\Delta A_{\rm X}$ 
vs. $T_{\rm c}$ relation we obtained ($+2.37\times 10^{-5}$~dex/K) is steeper 
in comparison with those already reported in the previous papers;
$+1.88 \times 10^{-5}$ (Tucci Maia et al. 2014),
$+0.98 \times 10^{-5}$ (Nissen et al. 2017), and
$+1.56 \times 10^{-5}$ (Tucci Maia et al. 2019).
This is because our $\Delta A$ values of the volatile CNO (especially O) tend 
to be lower than theirs (while those for the refractory elements are mostly 
in agreement), which makes the contrast between $\Delta A$(volatile) 
and $\Delta A$(refractory) more conspicuous in our result.
Yet, we had better be cautious about the significance of this discrepancy, 
since it is mainly due to the comparatively lower oxygen abundance of 
less reliability (especially for 16~Cyg~A; cf. Sect.~4.2.4)  
As a matter of fact, the slope becomes more gradual if O is excluded 
($+1.28 \times 10^{-5}$), which then becomes almost consistent with 
the results of other authors.

\section{Summary and conclusion}

The visual binary system 16~Cyg~A+B is of particular astrophysical interest, 
because a planetary companion is detected only in B but not in A, despite that
both are solar twins quite resembling each other.

It is important to clarify whether any chemical abundance differences exist 
between A and B (i.e., in the metallicity as well as in the relative abundance 
patterns), which may provide useful information regarding the impact of planet 
formation upon the host star.

A number of papers treating this issue have been published over the past several 
decades but with rather diversified results, reflecting the fact
that the differences (if any exist) are small and delicate.
I also carried out high-precision differential analyses on selected 
solar-analogue stars including 16~Cyg~A and B (Takeda 2005) and concluded that 
A and B have practically the same metallicity (within an uncertainty of $\pm 0.01$~dex).

However, several papers of other authors published thereafter reported results 
against this conclusion; they all arrived at a similar consequence that 16~Cyg~A 
is slightly more metal-rich than B by a few hundredths~dex (cf. Table~1). 
These recent studies also revealed that not only the difference in metallicity 
but also the different $T_{\rm c}$-dependent trend in elemental abundances exist 
between A and B.

Realising the necessity of revisiting this problem given this situation, I decided 
to conduct an intensive comparative analysis of the Sun, 16~Cyg~A, and 16~Cyg~B 
based on the public-domain high-dispersion spectra obtained with CFHT/ESPaDOnS.
Special attention was paid to the following points: 
(i) precisely establishing the differential atmospheric parameters/metallicity 
between A and B based on Fe~{\sc i} and Fe~{\sc ii} lines
(while clarifying the nature/uncertainty of solutions in the parameter space),  
(ii) deriving the abundances of various refractory elements (of high $T_{\rm c}$) 
from equivalent widths, and (iii) determining the abundances of CNO (volatile elements 
of low $T_{\rm c}$) by applying the spectrum-fitting technique to the line features 
of CH, NH, and OH molecules. 

The differential Fe abundances between 16~Cyg~A (A), 16~Cyg~B (B), and 
the Sun (S) turned out 
$\Delta A_{\rm Fe}^{\rm A-S}$ = +0.09, 
$\Delta A_{\rm Fe}^{\rm B-S}$ = +0.06, and
$\Delta A_{\rm Fe}^{\rm A-B}$ = +0.03 
(with an uncertainty of $\sim \pm 0.01$), which means that A is slightly more metal-rich 
than B by +0.03~dex. This lends support to the results of several recent work done 
in the past decade. Accordingly, I admit that the conclusion (almost equal metallicity
for A and B) once derived by myself (Takeda 2005) was incorrect, which was presumably 
due to inadequate observational $W_{\lambda}$ data used therein (e.g., inclusion of
lines in longer wavelength region where the quality of the spectra was insufficient).

The differential abundances ($\Delta A_{\rm X}$) of CNO (volatile elements of low 
$T_{\rm c}$) are comparatively lower than those of refractory elements of 
higher $T_{\rm c}$, leading to a positive slope in the $\Delta A_{\rm X}$ vs. $T_{\rm c}$ 
relation, and this holds for any of 
$\Delta A_{\rm X}^{\rm A-S}$,
$\Delta A_{\rm X}^{\rm B-S}$, and
$\Delta A_{\rm X}^{\rm A-B}$.

Although  this trend is qualitatively consistent with the recent studies of 
other authors, the slope derived from this investigation is steeper than theirs,
which is because the CNO abundances obtained from the lines of hydride molecules 
tend be somewhat lower than those derived (mostly from atomic lines) by others. 
We should note, however, that this is mainly attributed to the inclusion of less 
reliable O abundances (which are subject to larger errors). If O is excluded, 
the resulting slope of the linear-regression line becomes milder and is 
favourably compared with previous determinations.    

It may also be worth noting that $\Delta A_{\rm C}$ is larger than $\Delta A_{\rm N}$ 
and/or $\Delta A_{\rm O}$, which is not compatible with the global trend. 
If this is meaningful, some factor other than $T_{\rm c}$ might be involved 
in the CNO abundances.
     
Regarding the origin of abundance differences between 16~Cyg~A and B, much 
can not be said based on these results alone. Yet, some kind of a posteriori 
chemical pollution mechanism (possibly related to planet formation) is likely have 
acted in this binary system but differently on A and B, as argued by recent studies
(e.g., planet-engulf event having acted on A as speculated by Tucci Maia et al. 
2019 on the assumption that it might have hosted a planet in the past). 

\section*{Acknowledgements}

This study made use of the public-domain data provided by Canada--France--Hawaii
Telescope. This investigation is also based partly on the data collected at 
Subaru Telescope, which is operated by the National Astronomical Observatory of Japan.
This research has made use of the SIMBAD database, operated by CDS, 
Strasbourg, France. 

\section*{Data availability}

The basic data and results underlying this article are presented as 
the online supplementary material. Regarding the observational data used 
in this study, the reduced CFHT/ESPaDOnS spectra are downloadable from 
web site of the Canadian Astronomy Data Centre 
(https://www.cadc.hia.nrc.gc.ca/AdvancedSearch/), while the original 
Subaru HDS data are available at the SMOKA Science Archive site 
(https://smoka.nao.ac.jp/index.jsp).

\section*{Supporting information}

This article accompanies the following online materials.
\begin{itemize}
\item
{\bf readme.txt} 
\item
{\bf ewlines.dat} 
\item
{\bf lines\_OHregions.dat}
\item
{\bf relabunds\_AtoS.dat} 
\item
{\bf relabunds\_BtoS.dat} 
\item
{\bf relabunds\_AtoB.dat} 
\item
{\bf CNOabunds.dat} 
\end{itemize}

\end{document}